\documentclass[article]{elsarticle}
\usepackage{pifont}
\usepackage{natbib}
\usepackage{amssymb}
\usepackage{amsmath}
\usepackage{graphicx}
\usepackage{geometry}

\def\tfract#1/#2{{\textstyle{\raise0.8pt\hbox{$\scriptstyle#1$}\over%
\hbox{\lower0.8pt\hbox{$\scriptstyle#2$}}}}}

\def\radi2k{\tfract 1/{\sqrt {2k}} }
\def\der{\partial }
\def\downnormalfill{$\,\,\vrule depth4pt width0.4pt
\leaders\vrule depth 0pt height0.4pt\hfill\vrule depth4pt width0.4pt\,\,$}
\def\WT#1{\mathop{\vbox{\ialign{##\crcr\noalign{\kern3pt}
      \downnormalfill\crcr\noalign{\kern0.8pt\nointerlineskip}
      $\hfil\displaystyle{#1}\hfil$\crcr}}}\limits}

\def\be{\begin{equation}}
\def\ee{\end{equation}}
\def\bea{\begin{eqnarray}}
\def\eea{\end{eqnarray}}
\def\ba{\begin{array}{rcl}}
\def\ea{\end{array}}
\def\der{\partial}
\def\go{\leavevmode \raise.3ex\hbox{$\scriptscriptstyle \langle\!\langle\!  $}%
~\ignorespaces}

\def\gf{\relax \ifhmode \unskip~\else \leavevmode \fi \raise.3ex\hbox{$\! \scriptscriptstyle\rangle\!\rangle\, $}}
\catcode`@=11 
\font\sevensy=cmsy7
\catcode`@=12 
\newbox\novebox
\setbox\novebox=\hbox{\kern+.3em {$V_t$}%
\kern-0.79em\raise1.8ex\hbox{\sevensy {\char'016}}\kern+.3em}

\newbox\ottobox
\setbox\ottobox=\hbox{\kern+.3em {$V$}%
\kern-0.61em\raise1.8ex\hbox{\sevensy {\char'016}}\kern+.3em}
\def\vnt{\copy\ottobox}
\begin{document}

\pagestyle{empty}
\setcounter{page}{0}

\begin{center}
{\Large {\bf Path-integral invariants in abelian Chern-Simons theory}}
\\[1cm]
{\large E.~Guadagnini$^a$ and F.~Thuillier$^b$}

\end{center}

\vskip 0.7 truecm

{$^a$ \it Dipartimento di Fisica ``E. Fermi",  Universit\`a di Pisa,  Largo B. Pontecorvo 3, 56127 Pisa; and INFN, Italy.}

{$^b$ \it LAPTh, Universit\'e de Savoie, CNRS, Chemin de Bellevue, BP 110,  F-74941 Annecy-le-Vieux Cedex, France.}

\vspace{1cm}
\begin{center}
{\bf Abstract}
\end{center}

\noindent We consider the $U(1)$ Chern-Simons gauge theory defined in a general closed  oriented 3-manifold $M$; the   functional integration is used to compute the normalized partition function and the expectation values of the link holonomies. The nonperturbative path-integral is defined in the space of the gauge orbits of the connections which belong to the various  inequivalent $U(1)$ principal bundles over $M$; the different sectors of the configuration space  are labelled by the elements of the first homology group of $M$ and are characterized by appropriate background connections. The gauge orbits of flat connections, whose classification is also based on the  homology group, control the  extent  of the nonperturbative contributions to the mean values.  The  functional integration is achieved in any 3-manifold $M$,  and the corresponding path-integral invariants turn out to be  strictly related with the abelian Reshetikhin-Turaev surgery invariants.

\newpage

\section{Introduction}

In a recent article \cite{1} we have presented a path-integral  computation of the normalized partition function $Z_k(M)$ of the $U(1)$ Chern-Simons (CS) field theory \cite{2,3,4} defined in a  closed oriented 3-manifold $M$. It has been shown \cite{1} that, when the first homology group $H_1(M)$ is finite,  by means of the functional integration  one recovers ---in a nontrivial way--- the abelian  Reshethikin-Turaev \cite{5,6,7,8} surgery invariant.

The present article completes the construction  of the path-integral solution of the $U(1)$ quantum CS field theory initiated in Ref.~\cite{9}.    We extend the computation of $Z_k(M)$ to the general case in which the homology group of $M$ is not necessarily finite and may  contain nontrivial free (abelian) components. We give a detailed description of abelian gauge theories in topological nontrivial manifolds, and  the resulting extension of the gauge symmetry group is discussed. We classify the gauge orbits of flat connections; their role in the functional integration is determined.  The path-integral
  computation of both the perturbative and nonperturbative components  of the expectation values of the gauge holonomies associated with  oriented colored framed links  is illustrated.  The result of the functional integration is compared with the combinatorial invariants of  Reshetikhin-Turaev; it is found that  the path-integral invariants are  related with the abelian surgery invariants of Reshetikhin-Turaev by means of a nontrivial multiplicative factor which only depends on the torsion numbers and on the first Betti number of the manifold $M$.

A general outlook on the nonperturbative method ---which is used to achieve the complete functional integration of the observables for the abelian CS theory  in a general manifold $M$---  is contained in Section~2; the details are given in the remaining sections.  As in our previous articles \cite{1,9,10}, we use the  Deligne-Beilinson (DB) formalism \cite{11,12,13}  to deal with  the  $U(1)$ gauge fields;  the  functional integration amounts to  a sum over the inequivalent $U(1)$ principal bundles over $M$ supplemented by  an  integration over the gauge orbits of  the corresponding connections.   The essentials of the Deligne-Beilinson formalism are collected in the Appendix.  The structure of the configuration space is described in Section~3, where the path-integral normalization of the partition function  and of the reduced expectation values is also  introduced.  Section~4 contains a description of the  gauge orbits of $U(1)$ flat connections in the manifold $M$,  the classification of the different types of flat connections is based on the first homology group of $M$.  The functional integration is accomplished in Section~5, and the comparison of the path-integral invariants with the surgery Reshetikhin-Turaev invariants is contained in Section~6. Examples of computation of path-integral invariants in lens spaces are reported in Section~7. Section~8 contains the conclusions.

\section{Overview}

 The functional integration in the abelian CS field theory can be accomplished by means of the nonperturbative method developed in \cite{1,9,10}. In order to introduce progressively the main features of this method, let us first consider the case of  a homology sphere $M_0$, for which the first homology group $H_1(M_0)$ is trivial;  the 3-sphere $S^3$ and the Poincar\'e manifold  are examples of homology spheres.   Let us recall that the homology group of a manifold $M$  corresponds to the abelianization  \cite{14} of the fundamental group $\pi_1 (M)$; {\it i.e.} given a presentation of $\pi_1 (M)$ in terms of generators and relations,  a presentation of $H_1(M)$  can be obtained  by imposing the additional constraint that the generators of $\pi_1 (M)$ commute.

  The field variables of the $U(1)$ CS theory in $M_0$ are described by a 1-form $ A \in \Omega^1 (M_0) $ with components $  A = A_\mu(x) dx^\mu $, and the action is
\be
S [ A] =  2 \pi k \int_{M_0}  d^3 x \, \varepsilon^{\mu \nu \rho }\, A_\mu  \partial_\nu  A_\rho  =  2 \pi k \int_{M_0}  A \wedge d  A \; ,
\label{2.1}
\ee
where $k \not= 0$ denotes the real coupling constant of the model. The action is invariant under usual gauge transformations $ A_\mu (x) \rightarrow  A_\mu (x) + \der_\mu \xi (x)$. This means that the action can be understood as  a function of the gauge orbits.

\subsection {Generating functional} In order to define the expectation values $\langle  A_\mu(x)   A_\nu (y) \cdots  A_\lambda (z)\rangle $ of the products of fields,   one needs to introduce a gauge-fixing procedure because  the gauge field  $ A_\mu (x)$ is not gauge-invariant.  However, if one is interested in the correlation functions  $\langle F_{\mu \nu} (x) F_{\rho \sigma }(y) \cdots F_{\lambda \tau}(z) \rangle$ of the curvature   $F_{\mu \nu}(x)=  \der_\mu A_\nu(x) - \der_\nu A_\mu (x) $,  the  gauge-fixing  is not required. In facts, let us introduce a classical external source which is described by  a  1-form $ B = B_\mu (x) dx^\mu $; the integral
\be
   \int d  A \wedge  B  = \int    A \wedge   d B
   \label{2.2}
\ee
 is invariant under gauge transformations acting on $ A$ because the curvature $F = d A $ is gauge-invariant.  The generating functional ${\cal G} [  B ]  $ for the correlation functions of the curvature is defined by
\be
{\cal G} [B]  = \left  \langle e^{2 \pi i \int A \wedge d B } \right  \rangle \equiv    {\int DA \; e^{2 \pi i k \int A \wedge d A } \; e^{ 2 \pi i \int A \wedge d B } \over \int DA \; e^{2 \pi i k \int A \wedge d A }}  \; ,
\label{2.3}
\ee
indeed the coefficients of the Taylor expansion of ${\cal G} [B]$ in powers of $B$ coincide with the correlation functions of the curvature. Any configuration $A_\mu (x) $ can be written as
\be
A_\mu (x) = - {1 \over 2k} B_\mu (x) + \omega_\mu (x) \; ,
\label{2.4}
\ee
where $B_\mu (x)$ is fixed and   $\omega_\mu (x) $ can fluctuate. Since
\be
k \int_{M_0} A \wedge d A + \int_{M_0} A \wedge d B = k \int_{M_0} \omega \wedge d \omega - {1\over 4k} \int_{M_0} B \wedge B \; ,
\label{2.5}
\ee
and  the functional integration is invariant under translations, {\it i.e.} $DA = D \omega$, one finds
\be
\left  \langle e^{2 \pi i \int A \wedge d B } \right  \rangle  = e^{- (2 \pi i / 4 k ) \int B \wedge d B} {\int D\omega \; e^{2 \pi i k \int \omega \wedge d \omega } \over \int DA \; e^{2 \pi i k \int A \wedge d A }}
= e^{- (2 \pi i / 4 k ) \int B \wedge d B} \; .
\label{2.6}
\ee
So without the introduction of any gauge-fixing  ---and hence without the introduction of any metric in $M$--- the Feynman path-integral gives
\be
 {\cal G}[B] = \exp \left ( i {\cal G}_c [B] \right ) = \exp \left (  - {2 \pi i \over  4 k } \int_{M_0} B \wedge d B \right ) \; .
 \label{2.7}
\ee
The generating functional  of the connected correlation functions of the curvature ${\cal G}_c[B] $ formally coincides with the Chern-Simons action (\ref{2.1})  with the replacement $k \longrightarrow -1/4k$.

\medskip

\newdefinition{rmk}{Remark}
\begin{rmk}
The result (\ref{2.6})  can also be obtained  by means of the standard perturbation theory  with, for instance,  the BRST gauge-fixing procedure of the Landau gauge; in the case of the abelian CS theory, the method presented in Ref.\cite{15} can be used   in any homology sphere.  Expression (\ref{2.7}) is also a consequence of the Schwinger-Dyson equations. Indeed the only connected diagram entering ${\cal G}_c[B] $  is given by the two-point function of the curvature $\langle \varepsilon^{\mu \nu \rho} \der_\nu A_\rho (x) \varepsilon^{\lambda \sigma \tau } \der_\sigma A_\tau (y) \rangle = N^{-1}\int DA \, e^{iS[A]} \varepsilon^{\mu \nu \rho} \der_\nu A_\rho (x) \varepsilon^{\lambda \sigma \tau } \der_\sigma A_\tau (y)  $. Since $\varepsilon^{\mu \nu \rho} \der_\nu A_\rho (x)= (1/ 4 \pi k )\, \delta S[A] / \delta A_\mu (x) $, one finds
\bea
\langle \varepsilon^{\mu \nu \rho} \der_\nu A_\rho (x) \varepsilon^{\lambda \sigma \tau } \der_\sigma A_\tau (y) \rangle &=& (-i / 4 \pi k ) N^{-1}\int DA \, ( \delta e^{i S[A]} / \delta A_\mu (x)) \, \varepsilon^{\lambda \sigma \tau } \der_\sigma A_\tau (y) \nonumber \\
&=& (i / 4 \pi k ) N^{-1}\int DA \, e^{i S[A]} \varepsilon^{\lambda \sigma \tau } \, \delta [ \der_\sigma A_\tau (y)] / \delta A_\mu (x)\nonumber \\
&=& -\,  {i \over 4\pi k} \, \varepsilon^{\lambda \sigma \mu } \, {\der \over \der x^\sigma}  \delta^3 (x-y) \; ,
\label{2.8}
\eea
which is precisely the kernel appearing in  ${\cal G}_c[B] $.
 \end{rmk}

\begin{rmk}
Since the action (\ref{2.1}) and the source coupling (\ref{2.2}) are both invariants under gauge transformations $ A_\mu (x) \rightarrow  A_\mu (x) + \der_\mu \xi (x)$, in the computation of the expectation value (\ref{2.3}) the functional integration can be interpreted as an integration over  the gauge orbits.
\end{rmk}

 The generating functional (\ref{2.7}),  which  gives the solution of the abelian CS theory in $M_0$, depends on the smooth classical source $B_\mu (x)$.  In order to bring the topological content of $ {\cal G}[B]$  to light, it is convenient to consider the limit in which the source $B_\mu (x)$ has support of knots and links in the manifold $M_0$.

\medskip

\subsection {Knots and links} For each oriented knot $C\subset M_0$ one can introduce \cite{9,13,16,17} a de~Rham-Federer 2-current $j_C$ such that, for any 1-form $\omega $, one has   $\oint_C \omega = \int_{M_0} \omega \wedge j_C   $. Moreover, given a Seifert surface $\Sigma$ for $C$ (that verifies $\partial \Sigma = C$), the associated 1-current $\alpha_\Sigma $ satisfies $j_C = d \alpha_\Sigma $ and then    $\oint_C\omega = \int_{M_0} \omega \wedge j_C  = \int_{M_0} \omega \wedge d \alpha_\Sigma $. So, given the link $C_1 \cup C_2 \subset M_0$, the linking number of $C_1 $ and $C_2$ is given by $\ell k (C_1 , C_2) = \int_{M_0} j_{C_1} \wedge \alpha_{\Sigma_2} = \int_{M_0} \alpha_{\Sigma_1} \wedge j_{C_2} $ without the introduction of any regularization.

Let $L = C_1 \cup C_2 \cup \cdots \cup C_n \subset M_0$ be a oriented framed colored  link in which the knot $C_j$  is provided with the framing  $C_{j{\rm f}}$ and its color is specified by the real charge $q_j$. Let us introduce the  1-current $\alpha_L := \sum_j q_j \alpha_{\Sigma_j}$ where $C_j $ is the boundary of the surface $ \Sigma_j $. In the  $B \rightarrow \alpha_L$ limit, equation (\ref{2.6}) becomes \cite{9}
\bea
\left  \langle e^{2 \pi i \int A \wedge d \alpha_L } \right  \rangle &=&
\left  \langle    e^{2 \pi i \sum_{j=1}^n q_j \oint_{C_j} A  } \right \rangle   \equiv
\left  \langle W_L (A) \right \rangle \Big |_{M_0}   = \nonumber \\
&=& \exp \left (  - {2 \pi i \over  4 k } \int_{M_0} \alpha_L \wedge d \alpha_L \right )
 =   \exp \Bigl  ( - {2 \pi i  \over 4k}  \, \Lambda_{M_0} (L , L) \Bigr ) \; ,
 \label{2.9}
\eea
in which  the quadratic function $\Lambda_{M_0} (L , L) $ of the link $L$ is given by
\be
\Lambda_{M_0} (L , L) =  \sum_{i,j=1}^n q_i   q_j  \, \ell k (C_i , C_{j{\rm f}}) \Big |_{M_0} \; ,
\label{2.10}
\ee
where $\ell k (C_i , C_{j{\rm f}}) \big |_{M_0} $ denotes the linking number of $C_i $ and $C_{j{\rm f}}$ in $M_0$.  Note that, for integer values of the charges $q_i$, $\Lambda_{M_0} (L , L)$ takes integer values.   The $B \rightarrow \alpha_L$ limit can be taken after the path-integral computation or   directly before   the functional integration; in both cases expression (\ref{2.9}) is obtained.

\subsection {The complete solution}  When the abelian CS theory is defined  in a 3-manifold $M$ which is not a homology sphere, the formalism presented above needs to be significantly improved  in various aspects.

\medskip

\noindent (1) {\it Gauge symmetry.} The first issue is related with the gauge symmetry. We consider the CS gauge theory in which the fields are $U(1)$ connections on $M$; when $M$ is not a homology sphere, $U(1)$ gauge fields are no more described by 1-forms, one needs additional variables to characterize gauge connections.  Each connection can be described by a triplet of local field variables which are defined in the open sets of a good cover of $M$ and in their intersections.
The gauge orbits  of the $U(1)$ connections will be described by DB classes belonging to the space $H^1_D (M)$; a few basic definitions of the Deligne-Beilinson formalism can be found in the Appendix. In the DB approach ---as well as in any formalism in which the $U(1)$ gauge holonomies represent a complete set of observables--- the charges $q_j $ and the coupling constant $k$ must assume integer values.

\medskip

\noindent (2) {\it Configuration space.}  Each gauge connection  refers to a $U(1)$ principal  bundle over $M$ that may be nontrivial, and the space of the gauge orbits accordingly  admits a canonical decomposition into various disjoint sectors or fibres which can be labelled by the elements of  the first homology group $H_1(M)$   of $M$.  As far as the functional integration is concerned,   the important point is that all the gauge orbits of a given fibre  can be obtained by adding 1-forms (modulo 1-forms of integral periods which corresponds to gauge transformations)  to a chosen fixed orbit,  that can be interpreted as an origin element of the fibre and plays the role of a background gauge configuration. For each element of $H_1(M)$ one has an appropriate background connection.  Thus the functional integration  in each fibre consists of a  path integration over 1-form variables  in the presence of a (in general non-trivial) gauge background which characterizes the fibre. Then,  in the entire functional integration, one has to sum over all the backgrounds.

Each path-integral with fixed background can be normalized with respect to the functional integration in presence of the trivial background of the vanishing connection; in this way one can give a meaningful definition \cite{1}  the   partition function of the CS theory.

Since the homology group of a homology sphere is trivial,  in the case of a homology sphere  the space of gauge connections consists of a single fibre ---the set of 1-forms modulo gauge transformations---  and the corresponding origin, or  background field, can be taken to be the null connection; so   one recovers the circumstances described in \S~2.1 and \S~2.2.

\medskip

\noindent (3) {\it Chern-Simons action.}   In the presence of a nontrivial $U(1)$ principal bundle, the dependence of  the CS action on the gauge orbits of the corresponding connections is not  given by expression (\ref{2.1}); one needs to improve the definition of the CS action  so that $U(1)$ gauge invariance is maintained. In the DB formalism, the gauge orbits of $U(1)$ connections are described by the so-called DB classes; for each class $A \in H^1_D(M)$ the abelian CS action is given by
\be
S[A] = 2 \pi k \int_M A * A \; ,
\label{2.11}
\ee
where $A * A$ denotes the DB product \cite{13} of $A$ with $A$, which represents a generalization of the lagrangian appearing in equation (\ref{2.1}); details on this point can be found in the Appendix.

\medskip

\noindent (4) {\it Generalized currents.} When the homology class of a knot $C \subset M$  is not trivial,  there is no Seifert surface $\Sigma$ with boundary $\partial \Sigma = C$; consequently one cannot define a  1-current $\alpha_\Sigma $ associated with $C$.  Nevertheless, the standard de~Rham-Federer theory of currents  admits a  generalization \cite{9}  which is based on  appropriate distributional DB classes. This means that, for any link $L \subset M$, one can find a distributional DB class $\eta_L$ such that  the abelian holonomy associated with $L$ can be written as
\be
\exp \left ( 2 \pi i \oint_L A \right )  \; \longrightarrow \; \exp  \left ( 2 \pi i \int_M A * \eta_L \right ) = \hbox{ holonomy} \; .
\label{2.12}
\ee
In the case of a homology sphere, expression (\ref{2.12}) coincides with the gauge invariant coupling $\int A \wedge d \alpha_L$ appearing in equation (\ref{2.9}),  $\eta_L$ being  given by $\alpha_L$.

\medskip

\noindent (5) {\it Nonperturbative  functional integration.} When  trying to compute the  expectation values of the holonomies, one encounters the following path-integral
\be
\int DA \; e^{2 \pi i  \int_M  \left (  k \, A*A +  A * \eta_L  \right ) } \; .
\label{2.13}
\ee
In order to achieve the  functional integration over the DB classes  by using the nonperturbative method illustrated above, one would like to introduce a change of variables which is similar to the change of variables illustrated in equation (\ref{2.4}), namely
\be
\hbox {`` } A = - {1\over 2 k}  \, \eta_L + A^\prime \hbox{ "} \; ,
\label{2.14}
\ee
where $A^\prime $ denotes the fluctuating  variable. Unfortunately, as it stands equation (\ref{2.14})  is not coherent because the product of the rational number $(1 / 2k ) \not= 1$ with the DB class $\eta_L$ is not a DB class in general; in fact the abelian group $H^1_D(M)$ is not a linear space over the field ${\mathbb R}$ buth rather over ${\mathbb Z}$,  and   the naive use of  equation (\ref{2.14}) would   spoil gauge invariance. In order to solve this problem one needs to distinguish DB classes ---together with their  local representatives 1-forms--- from the 1-forms globally defined in $M$. It turns out that

 \begin{enumerate}[(i)]

\item  when the homology class $[L]$ of $L$ is trivial, one can define [9] a class $\eta^\prime_L $ such that $\eta^\prime_L + \eta^\prime_L + \cdots + \eta^\prime_L = (2k) \, \eta^\prime_L = \eta_L$ and, as it will be shown  in the Section~5, this solves the problem;

\item when the nontrivial element $[L]$  belongs to the torsion component of $H_1(M)$, one can always find a integer $p$ that trivializes the homology, $p [L] = 0$,  and then one can proceed  in a way  which is rather similar to the method adopted in case (i);

\item the real obstruction that prevents the introduction  of a change of variables of the type (\ref{2.14}) is found when $[L]$ has a nontrivial  component which belongs to the freely generated  subgroup of $H_1(M)$. But in this case there is really no need to change variables ---as indicated in equation (\ref{2.14})---  because  the direct functional integration over the zero modes gives a vanishing  expectation value  to the holomomy.

\end{enumerate}

\medskip

\noindent (6) {\it Flat connections.}  The nontriviality of the homology group $H_1(M)$ also implies the existence of gauge orbits of flat connections which have an important  role in the functional integration. On the one hand, the  flat connections which are related with the torsion component of the homology control the extent of the nonperturbative effects in the mean values and, on the other hand, the flat connections which are induced by the (abelian) freely generated component of the homology   implement the cancellation mechanism mentioned in point (iii) in the functional integration.

One eventually  produces a complete nonperturbative functional integration of the partition function and of the expectation values of the observables. So the abelian CS model is a particular example of a significant gauge quantum field theory that can be defined in a general oriented 3-manifold $M$,  the orbits  space of gauge connections is nontrivially structured according to the various inequivalent $U(1)$ principal bundles over $M$, the topology of the manifold $M$ is revealed by the presence of flat connections that give rise to nonperturbative contributions to the observables, and  one gets a full  achievement of   the path-integral computation.

\section{The quantum abelian Chern-Simons gauge theory}

Let the atlas ${\cal U} = \{ {\cal U}_a \}$ be a good cover of the closed oriented 3-manifold $M$; a $U(1)$ gauge connection $\cal A$ on $M$ can be described by a triplet of local variables
\be
{\cal A} = \{ v_a, \lambda_{a b} , n_{a b c} \} \; ,
\label{3.1}
\ee
where the $v_a$'s are 1-forms in the open sets ${\cal U}_a$, the $\lambda_{a b }$'s represent 0-forms (functions) in the intersections  ${\cal U}_a  \cap {\cal U}_b $ and the $n_{abc}$'s are integers defined in the intersections ${\cal U}_a \cap {\cal U}_b \cap {\cal U}_c $.
 The functions $\lambda_{ab}$  codify the gauge ambiguity $ v_b - v_a = d \lambda_{a b} $ in the intersection ${\cal U}_a \cap {\cal U}_b $. Similarly, the integers $n_{abc}$ ensure the consistency condition $ \lambda_{ b c } - \lambda_{a c} + \lambda_{a b } = n_{a b c} $
 that  the 0-forms $\lambda_{ab}$ must satisfy in the intersections ${\cal U}_a \cap {\cal U}_b \cap {\cal U}_c $.  The connection which is associated with a 1-form $\omega $ globally defined in $M$,   $\omega \in \Omega^1(M) $,  has components $ \{ \omega_a , 0 , 0 \}$, where $\omega_a$ is the restriction of $\omega $ in ${\cal U}_a$.

An element $\chi$ of $\Omega^1_{\mathbb Z} (M)$ ---which is called a form of integral period---  is a closed 1-form on $M$ such that, for any knot $C \subset M$, one has $\oint_C \chi = n \in {\mathbb Z}$.   Let us assume    that a complete set of observables is given by the set of holonomies $\{ \exp \left ( 2 \pi i \oint_L {\cal A} \right ) \}$  associated with links  $L \subset M$. Then  the connections $\cal A$ and ${\cal A} + \chi $ with $\chi \in \Omega^1_{\mathbb Z} (M) $ are gauge equivalent because there is no observable that can distinguish them. Consequenlty the space  $\Omega^1_{\mathbb Z} (M)$ of closed forms with integral periods corresponds to the set of gauge transformations.
The gauge orbit  $A$ of a given  connection $\cal A$ is the equivalence class of connections $\{ {\cal A} + \chi \} $ with varying $\chi \in \Omega^1_{\mathbb Z} (M) $.  Each gauge orbit can be represented by one generic element of the class, and the notation
$$
A \leftrightarrow \{ v_a, \lambda_{a b} , n_{a b c} \}
$$
means that the class $A$ can be represented by the connection ${\cal A}=  \{ v_a, \lambda_{a b} , n_{a b c} \}$.

The configuration space of a $U(1)$ gauge theory is given by the set of equivalence classes  of $U(1)$ gauge connections on $M$ modulo  gauge transformations, and can be identified  with the cohomology space $H^1_D(M)$ of the Deligne-Beilinson  classes. This space  admits a canonical  fibration over the first homology group $H_1(M)$  which is induced by the exact sequence
\be
0 \rightarrow \Omega^1(M) / \Omega^1_{\mathbb Z} (M) \rightarrow H^1_D(M) \rightarrow  H_1(M) \rightarrow 0 \; .
\label{3.2}
\ee
Hence the  space $H^1_D(M)$  can be interpreted as a disconnected  affine space whose connected components are indexed by  the elements of the  homology group of $M$.  The  1-forms modulo closed forms with integral periods  ---{\it i.e.} the elements of $\Omega^1(M) / \Omega^1_{\mathbb Z} (M)$--- act as translations on each connected component.   A  picture of $H^1_D(M)$ is shown in Figure~3.1; the different fibres match the inequivalent $U(1)$ principal bundles over $M$ and, for fixed principal bundle, the elements of each fibre  describe  the gauge orbits of the corresponding connections.

\vskip 0.5 truecm
\centerline {\includegraphics[width=1.99 in]{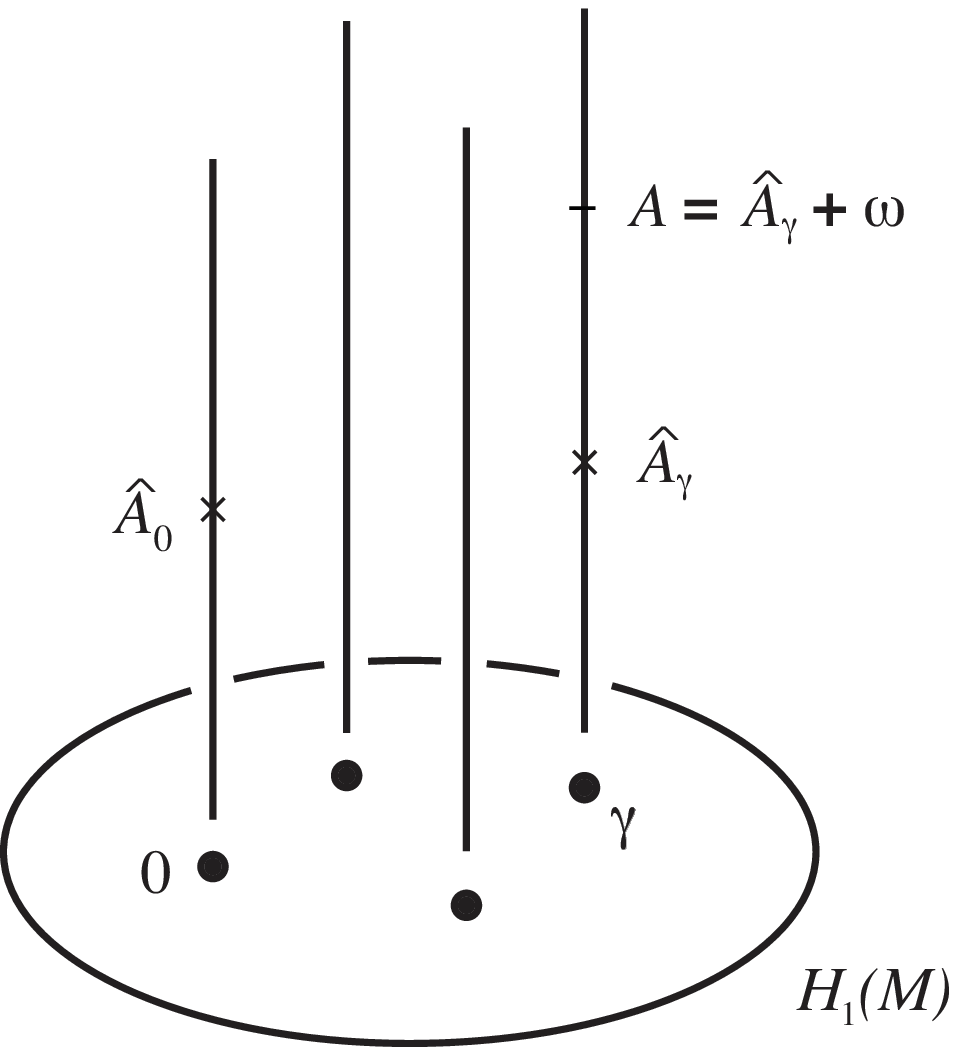}}
\vskip 0.3 truecm
\centerline {{Figure 3.1.} {Fibration  of $H_D^1 (M)$ over $H_1(M)$.}}

\vskip 0.6 truecm

\noindent Each class $A\in H_D^1(M)$ which belongs to the fibre over the element $\gamma \in H_1(M)$ can be written as
\be
A = \widehat A_\gamma + \omega \; ,
\label{3.3}
\ee
where $\widehat A_\gamma$ represents a specified  origin in the fibre  and $\omega \in \Omega^1(M) / \Omega^1_{\mathbb Z} (M)$. The choice of the  class $ \widehat A_\gamma $ for each element $\gamma \in H_1(M)$ is not unique.  One can take $\widehat A_0 =0 $ as the canonical origin of the fibre over the trivial element of $H_1(M)$.

The abelian CS field theory is a $U(1)$ gauge theory with action $S[A]$  given by the integral on $M$ of the DB product $A* A$,
$ S[A] = 2 \pi k \int_M  A * A $,
where $k$ is the (nonvanishing)  integer coupling constant of the theory. A modification of the orientation of $M$ is equivalent to a change of the sign of $k$, so one can assume $k>0$.  The properties of the DB $*$-product  have been discussed for instance in Ref.\cite{13};  the explicit decomposition of $S[A]$ in terms of the field components can also be found in the Appendix.  The  functional integration is modeled \cite{1,9} on the  configuration space's structure. According to  equation (\ref{3.3}),  the whole path-integral is assumed to be given by
\be
\int DA \, e^{i S[A]} = \sum_{\gamma \in H_1(M)} \int D \omega \, e^{i S[\widehat A_\gamma + \omega ]} \; .
\label{3.4}
\ee
Since the CS action is a quadratic function of $A$,  the result of the  functional integration does not depend on the  particular choice of the origins $\widehat A_\gamma$.   Then one has to fix the overall normalization because only the ratios  of functional integrations  can be well defined.  A natural possibility \cite{1} is to choose the overall normalization to be given by  the integral over the  gauge orbits of the connections of the trivial $U(1)$ principal bundle over $M$, that is the integral over the 1-forms globally defined in $M$ modulo closed forms of integral periods.

\newdefinition{dfn}{Definition}
\begin{dfn}
{\it For each function $X(A)$  of the DB classes, the corresponding reduced expectation value $ \langle\!  \langle  X(A) \rangle \! \rangle  \big |_{M}$ is defined by  }
 \bea
 \langle\!  \langle  X(A) \rangle \! \rangle \Big |_{M}  &\equiv&    {\sum_{\gamma \in H_1(M)} { \int D \omega \, e^{i S[\widehat A_\gamma + \omega]} \, X ( \widehat  A_\gamma + \omega )}  \over  \int D \omega \, e^{i S[ \widehat A_0  + \omega]} } \nonumber \\
 &=& \sum_{\gamma \in H_1(M)} {  \int D \omega \, e^{i S[\widehat A_\gamma + \omega]} \, X ( \widehat  A_\gamma + \omega ) \over  \int D \omega \, e^{i S[ \omega]} }  \; .
\label{3.5}
\eea

\end{dfn}

\noindent When $X(A) = 1$,  one obtains the normalized partition function
\be
Z_k(M) \equiv \langle\!  \langle  1 \rangle \! \rangle \Big |_{M} =  \sum_{\gamma \in H_1(M)} { \int D \omega \, e^{i S[\widehat A_\gamma + \omega]}  \over  \int D \omega \, e^{i S[ \omega]} } \; .
\label{3.6}
\ee

\begin{rmk}
Note that the standard expectation values $\langle X(A)\rangle \big |_M$ are defined by
\be
 \langle  X(A)  \rangle \Big |_{M}  \equiv    {\sum_{\gamma \in H_1(M)} { \int D \omega \, e^{i S[\widehat A_\gamma + \omega]} \, X ( \widehat  A_\gamma + \omega )}  \over  \sum_{\gamma \in H_1(M)} \int D \omega \, e^{i S[ \widehat A_\gamma  + \omega]} } \; ,
 \label{3.7}
 \ee
 and can be expressed as
\be
  \langle  X(A)  \rangle \Big |_{M}  =  {  \langle\!  \langle  X(A) \rangle \! \rangle \Big |_{M}  \over  Z_k(M) }  \; .
\label{3.8}
\ee
The introduction of the reduced expectation values is useful because  it may happen that  $Z_k (M) $ vanishes and expression (\ref{3.7}) may formally diverge, whereas $\langle\!  \langle  X(A) \rangle \! \rangle \big |_{M} $ is always well defined.  By definition, for any homology sphere $M_0$ one has $Z_k (M_0) = 1$ because $H_1(M_0) =0$, and then in this case $\langle\!  \langle  X(A) \rangle \! \rangle \big |_{M_0} = \langle  X(A) \rangle \big |_{M_0}$.

\end{rmk}

Equation (\ref{3.5}) shows that the whole functional integration is given by a sum of  ordinary path-integrals over 1-forms $\omega $ in the presence of varying   background gauge configurations $\{ \widehat A_\gamma \}$; the  background fields $\{ \widehat A_\gamma \}$ characterize  the inequivalent   $U(1)$ principal bundles over $M$ and are labelled by the elements of the homology group of $M$.

 For each oriented knot $C\subset M$, the associated holonomy $W_C : H^1_D(M) \rightarrow U(1)$   is a function of $A$ which is denoted by $ W_C (A)= \exp  \left (  2 \pi i  \oint_C A\right ) $. The precise definition of the holonomy  $W_C (A)$ and its dependence on the fields components is discussed in the Appendix.

The holonomy $W_C(A)$   is an element of the gauge group $U(1)$;  in the irreducible $U(1)$ representation which is  labelled by $q \in {\mathbb Z} $, the holonomy $W_C(A)$ is represented by  $\exp  \left (  2 \pi i q \oint_C A\right )$.   Thus we consider  oriented colored knots in which the color of each knot is specified precisely by the integer value of a charge $q$.

In computing the expectation value $\langle\!  \langle  W_C \rangle \! \rangle \big |_{M}$ one finds  ambiguities because the expectation values of products of fields at the same point are not well defined. This is a standard feature of quantum field theory; differently from the products of classical fields at the same point ---that are well defined--- the path-integral mean values of the products of fields at the same points are  not well defined in general.   These ambiguities in $\langle\!  \langle  W_C \rangle \! \rangle \big |_{M}$ are completely removed  \cite{1,9} by introducing a  framing \cite{14} for each knot and by taking  the appropriate limit \cite{18} ---in order to define the mean value of the product of fields at coincident points--- according to the framing that has been chosen. As a result, at the quantum level, holonomies are really well defined for framed knots or for bands.   Given a framed oriented colored knot $C\subset M $, the corresponding expectation value $\langle\!  \langle  W_C \rangle \! \rangle \big |_{M}$ is well defined.

Consider a  framed oriented colored link $L  = C_1 \cup C_2 \cup \cdots \cup C_n \subset M$,    in which the color of the component $C_j$ is specified by the integer charge $q_j$ (with $j=1,2,..., n$); the gauge holonomy $W_L : A \rightarrow W_L(A)$ is just the product of the holonomies of the  single components
\be
  W_L(A) = e^{2 \pi i \oint_L A} \equiv  e^{   2 \pi i  q_1 \oint_{C_{1} }\! \! A } \, e^{   2 \pi i  q_2 \oint_{C_{2}} \! \! A } \cdots e^{   2 \pi i  q_n \oint_{C_{n}} \! \! A }   \; .
 \label{3.9}
 \ee
The expectation values $ \langle\!  \langle  W_L  \rangle \! \rangle \big |_{M} $ together with the partition function $Z_k (M)$ are the basic observables we shall consider.

\begin{rmk}
 The charge $q$ is quantized because it describes the irreducible representations of the gauge group $U(1)$.  Then  the group of  gauge transformations which do not modify the  value of the holonomies ---which are associated with colored links--- is given precisely by the set of closed 1-forms of integral periods. That is why the DB formalism is particularly convenient for  the description of  gauge theories with gauge group $U(1)$.   If a link component has charge $q=0$, this link component can simply be eliminated. If the oriented knot $C$ has charge $q$, a change of the orientation of $C$ is equivalent  to the replacement $q \rightarrow - q$.   The DB formalism also necessitates an integer coupling constant $k$.   For  fixed integer $k$,   the expectation values $ \langle\!  \langle  W_L  \rangle \! \rangle \big |_{M} $ are invariant  under the substitution $q_j \rightarrow q_j + 2k$ where $q_j$ is the charge carried by  a generic link component. This can easily be verified for homology spheres, see equation (\ref{2.10}), but has really a general validity \cite{9}.   Consequently one can impose that the charge $q$ of each knot  takes the values  $\{ 0,1,2,..., 2k -1 \}$; {\it i.e.} the color space coincides with the set of the residue classes of integers mod $2k$.
\end{rmk}

\begin{rmk}
 At the classical level,  the holonomy $\exp  \left (  2 \pi i q \oint_C A\right )$ ---for the oriented knot $C \subset M$ and integer charge $q > 1$--- can be interpreted as the holonomy associated with the path  $q \, C$,  in which the integral of $A$ covers  $q$ times the  knot $C$.
At the quantum level  the charge variables of the knots ---which refer to the color space---  admit a purely topological interpretation  based on satellites \cite{9,18} and on the band connected sums \cite{1,19, 20} of knots.
 \end{rmk}

\section {Homology and flat connections}

The homology group $H_1(M)$ of the 3-manifold $M$ is an abelian finitely generated group; it can be decomposed as
\be
H_1(M) = F(M) \oplus T(M) \; ,
\label{4.1}
\ee
where $F(M)$ is the so-called  freely generated   component
\be
F(M) = \underbrace{{\mathbb Z} \oplus {\mathbb Z} \oplus \cdots \oplus {\mathbb Z}}_B\; ,
\label{4.2}
\ee
with $ B \in {\mathbb N}$ commuting generators, and $T(M)$ denotes the torsion component
\be
T(M) = {\mathbb Z}_{p_1} \oplus {\mathbb Z}_{p_2} \oplus \cdots \oplus {\mathbb Z}_{p_N} \; ,  \label{4.3}
\ee
in which the integer  torsion numbers $\{ p_1 , p_2 ,..., p_N\}$ satisfy the requirement that $p_i$ divides $p_{i+1}$ (with $p_1 > 1$), and ${\mathbb Z}_p \equiv {\mathbb Z} / p  {\mathbb Z}$.  Let $\{ g_1 ,..., g_B \} $ and $\{ h_1 ,..., h_N \} $ denote the generators of $F(M)$ and $T(M)$ respectively; all generators commute and the generator $h_i$, with fixed $i=1,2,...,N$,  satisfies $ p_i h_i =0 $.

 The gauge orbits of $U(1)$ flat  connections in the manifold  $M$ are determined by the homology group $H_1(M)$. The two independent  components $F(M)$ and $T(M)$ of $H_1(M)$ correspond to two different kinds of flat connections.

To each element  $\gamma \in T(M)$ is associated the gauge orbit $A^0_\gamma$ of a flat connection. Since the de~Rham cohomology does not detect torsion [21], the gauge orbits $A^0_\gamma $  with $\gamma \in T(M)$  cannot be described by  1-forms; in fact,  the class $A^0_\gamma $ can be represented by the connection
\be
A^0_\gamma \leftrightarrow \{ 0 , \Lambda_{ab}(\gamma ) , N_{abc} (\gamma ) \} \; ,
\label{4.4}
\ee
where the first (1-form) component is vanishing, $\Lambda_{ab}(\gamma ) $ are rational numbers and $N_{abc} (\gamma )$ are necessarily nontrivial if $\gamma $ is not  trivial.   The curvature associated with $A^0_\gamma$ is vanishing, $dA^0_\gamma  = 0$, because the first  component of the representative connection (\ref{4.4}) is vanishing. An explicit construction of the class (\ref{4.4}) can be found in Ref.\cite{1}. Clearly, the gauge orbit $A_0^0$ can be represented by the vanishing connection $\{ 0 , 0 , 0 \}$. The classes (\ref{4.4}) can be taken as canonical origins for the fibres of $H^1_D(M)$ over $H_1(M)$ which are labelled by the elements of the torsion group $T(M)$.

To each generator $g_j$ (with $j =1,2,..., B$) of the freely generated subgroup $F(M)$  corresponds  a normalized zero mode $\beta^j \in \Omega^1(M)$;  $\beta^j $ is a closed 1-form  which is not exact
\be
d \beta^j =0 \quad , \quad \beta^j \not= d \xi_j \quad , \quad  \forall j = 1,2,...,B  \; ,
\label{4.5}
\ee
thus $\beta^j $ belongs to the first de~Rham cohomology space $H_{dR}^1(M)$. In facts the dimension of the linear space $H_{dR}^1(M)$ ---or the first Betti number---  is given precisely by $B$.  Zero modes can be normalized so that, if  the knot $C_{g_j} \subset M$ represents the generator $g_j$,
\be
\oint_{C_{g_j}} \beta^i = \delta^{\, i }_j \; ,
\label{4.6}
\ee
and, if the homology class of a knot $C\subset M$  has no components in $F(M)$, one has
\be
\oint_C \beta^j = 0 \; .
\label{4.7}
\ee

\begin{rmk}
For each mode $\beta^j$, let us consider the class $[\beta_j ]$ of 1-forms $\{ \beta^j + d \xi_j \}$ with varying $\xi_j \in \Omega^0(M)$;  one can represent this class $[\beta_j]$ by a specific distributional configuration ---or de~Rham-Federer current--- that can be denoted by $\widetilde \beta^j$. The 1-current $\widetilde \beta^j$ has support  on a closed oriented  surface $\Sigma_j $ that does not bound a 3-dimensional region of $M$ and thus $\Sigma_j $ represents an element of the second homology group $H_2(M)$. Indeed the group $H_2(M)$ is independent of torsion and it is only related with $F(M)$. More precisely, for each generator $g_i$ of $F(M)$ (with $i=1,2,..., B$) one can find a closed oriented surface $\Sigma_i \subset M$ which represents a generator of $H_2(M)$ such that the oriented  intersection of $\Sigma_i $ with $C_{g_j}$ is given precisely by $\delta^{\, i}_j$. Thus the 1-currents $\widetilde \beta^j$ with support on $\Sigma_j$ give an explicit distributional realization \cite{9} of the normalized zero modes     satisfying equations (\ref{4.6}) and (\ref{4.7}).
\end{rmk}

Let us now consider the gauge orbits of flat connections that are determined by the zero modes. For each zero mode $\beta^j$ one can introduce a set of DB classes $\omega^0 (\theta_j ) \in \Omega^1(M) / \Omega^1_{\mathbb Z}(M)$ which can be represented by
\be
\omega^0 (\theta_j) \leftrightarrow \{ \theta_j \beta^j_a , 0 , 0 \}  \; ,
\label{4.8}
\ee
where $\beta^j_a$ is the restriction of $\beta^j$ on ${\cal U}_a$ and the real parameter $  \theta_j  $ is  the amplitude of the mode $\beta^j$ in the class $\omega^0 (\theta_j)$. Since in each gauge orbit one needs to factorize the action of gauge transformations defined by closed 1-forms with integral periods, $\omega^0 (\theta_j )\rightarrow \omega^0 (\theta_j) + \chi $ with $\chi \in \Omega^1_{\mathbb Z}(M)$, the amplitude $\theta_j$ must take values in the circle $S^1$ which is given by the interval $I = [0, 1 ] $ with identified boundaries; that is $0 <  \theta_j \leq 1$.  The classes $\omega^0 (\theta_j )$ describe a set of gauge orbits of flat connections because, for any fixed value of the amplitude $\theta_j $,  one has  $d \omega^0 (\theta_j) \leftrightarrow \{ \theta_j \, d \beta_a^j = 0, 0 , 0 \} = 0$.

\begin{dfn}
{\it The zero modes, which are associated with the subgroup $F(M)$ of the homology, determine a set of gauge orbits of flat connections $\omega^0 (\theta) \in   \Omega^1(M) / \Omega^1_{\mathbb Z} (M)$ given by}
 \be
 \omega^0 (\theta )\, \leftrightarrow \{ \theta_1 \beta^1_a + \theta_2 \beta^2_a+ \cdots + \theta_B \beta^B_a , 0, 0 \} \; ,
\label{4.9}
\ee
 {\it in which $\beta^j_a$ is the restriction of $\beta^j$ on $\, {\cal U}_a $ and the real parameters  $\{ \theta_j \}$ satisfy $ 0 < \theta_j \leq 1$ for $j=1,2,..., B$. }

 \end{dfn}

\noindent  Therefore a generic element $\omega \in \Omega^1(M) / \Omega^1_{\mathbb Z} (M)$ can be decomposed as
 \be
\omega = \omega^0 (\theta ) + \widetilde \omega \; ,
\label{4.10}
\ee
where   $\, \widetilde \omega $ denotes what remains of the $\omega $ variables after the exclusion from $\Omega^1(M) / \Omega^1_{\mathbb Z} (M)$ of the  gauge orbits $ \omega^0 (\theta ) $, and the functional integration takes the form
\be
\int D \omega \; F [\omega ] =  \int_0^1d\theta_1 \int_0^1d\theta_2 \cdots \int_0^1 d \theta_B  \int D \widetilde \omega \; F[\omega^0(\theta ) + \widetilde \omega \, ] \; .
\label{4.11}
\ee

To sum up, the map of the gauge orbits of flat connections is given by
$$
H_1(M) {\buildrel \hbox{flat} \over {\hbox to 1.1 truecm {\rightarrowfill}  }}   \left \{ \begin{array} {l@{}l}
T(M)\rightarrow A^0_\gamma  &
 \hbox{ ~canonical origins  for the  fibres over } \gamma \in T(M) ; \\
F(M) \rightarrow \omega^0(\theta )  & \hbox { ~zero modes contributions to } \Omega^1(M) / \Omega^1_{\mathbb Z} (M) \; .
\end{array} \right.
$$

Finally, let us recall that  the holonomies   $ \{ \exp  \left (  2 \pi i  \oint_C {\cal A}_0\right ) \} $ ---which are associated with the knots $\{ C \subset M \} $--- of a flat connection ${\cal A}_0$   give a $U(1)$ representation of the fundamental group of $M$, which coincides with a $U(1)$ representation of $H_1(M)$ because the gauge group is abelian; the gauge orbit of a flat connection  is completely specified by this representation.

 The representation space ${\cal R}_\infty$ of  a free component ${\mathbb Z} \subset F(M)$ is the set of the possible values of the holonomy which are associated with a generator of ${\mathbb Z} \subset F(M)$; since ${\mathbb Z}$ is freely generated, one has ${\cal R}_\infty = U(1)$. For each zero mode $\beta^j$, with $j=1,2,...,B$,  let us consider the  $U(1)$ representation $\rho^{(j)} : H_1(M) \rightarrow U(1)$ which is defined by the holonomies of the flat connection ${\cal A}_0 = \theta_j \beta^j $ (no sum over $j$); equations (\ref{4.6}) and (\ref{4.7}) imply
 \be
 \begin{array}{l@{\; \; : \; \; }l}
 \rho^{(j)} & g_j \mapsto  e^{2 \pi i \theta_j} \; , \\
 \rho^{(j)} & g_i \mapsto  1 \quad , \quad \hbox{for }  i\not= j \; , \\
  \rho^{(j)} & h_i \mapsto 1 \; .
  \end{array}
  \label{4.12}
 \ee
 By varying $\theta_j $ in the circle $S^1$ given by $0 < \theta_j \leq 1 $, in equation (\ref{4.12}) one finds  the representation space ${\cal R}_\infty$.

 The representation space ${\cal R}_p$ of  a subgroup ${\mathbb Z}_p \subset T(M) $ ---which is given by the possible values of the holonomy of a generator of ${\mathbb Z}_p$--- coincides with the set of the $p$-th roots of unity, ${\cal R}_p= \{ \zeta^0 , \zeta^1 , \zeta^2 , ... , \zeta^{p-1} \}$ where $\zeta = e^{2\pi i / p}$. The representations $H_1(M) \rightarrow U(1)$ defined by the holonomies of the origins classes $A^0_\gamma$ (with $\gamma \in T(M)$) of equation (\ref{4.4}) depend on the manifold $M$. A few examples will be presented in Section~7.

\section{Functional integration}

This section contains the details of the functional integration for the partition function and for the abelian CS observables in a general manifold $M$.

\subsection{Opening}
 Given a framed oriented colored link $L  = C_1 \cup C_2 \cup \cdots \cup C_n \subset M$,    where the component $C_j$ has charge $q_j$ (with $j=1,2,..., n$), one can introduce \cite{9} a distributional DB class $\eta_L $ such that
the gauge holonomy $W_L(A)$ can be written as
\be
W_L(A) = \exp \left ( 2 \pi i \int_M A * \eta_L \right ) \; .
\label {5.1}
\ee
One can put
\be
\eta_L = \sum_{j=1}^n q_j \eta_{C_j} \; ,
\label{5.2}
\ee
in which the class $\eta_{C_j}$ can be represented by
\be
\eta_{C_j} \leftrightarrow \{ \alpha_a (C_j), \Lambda_{ab}(C_j) , N_{abc} (C_j)\} \; ,
\label{5.3}
\ee
where $\alpha_a (C_j)$ is a de~Rham-Federer 1-current defined in the open chart ${\cal U}_a$ such that $d \alpha_a (C_j)$ has support on the restriction of $C_j$ in ${\cal U}_a$. If the knot $C_j$ has trivial homology, then  $\alpha_a (C_j)$ can be taken to be the restriction in  ${\cal U}_a$ of a current $\alpha_{\Sigma_j}$ ---globally defined in $M$--- with support on a Seifert surface $\Sigma_j$ of $C_j$, and in this case the components  $\Lambda_{ab}(C_j) $ and $ N_{abc} (C_j)$ are trivial. If $C_j$ has nontrivial homology,  $\alpha_a (C_j)$ is no more equal to the restriction of a  globally defined 1-current and the components $\Lambda_{ab}(C_j) $ and $ N_{abc} (C_j)$ are necessarily nontrivial.

The homology class $[L]\in  H_1(M)$ of the colored link $L\subset M$ is defined to be the weighted sum ---weighted with respect to the values of the color charges--- of the homology classes of the link components
\be
[L] \equiv \sum_{i=1}^n q_i [C_i] =  [L]_F + [L]_T  \; ,
\label{5.4}
\ee
where
\be
[L]_F = \sum_{j=1}^B a^j_L \, g_j \quad , \quad [L]_T = \sum_{i=1}^N b_L^i \, h_i  \; ,
\label{5.5}
\ee
for certain  integers $ \{ a_L^j \} $ and $\{ b_L^i \}$.  There are no restrictions on the values taken by the integers $ \{ a_L^j \} $; whereas  the possible values of the integer $b_L^i$, for fixed $i$, belong to the residue class  of integers mod~$p_i$, because $p_i h_i =0$.

In order to compute the reduced expectation value
\be
 \langle\!  \langle  W_L(A) \rangle \! \rangle \Big |_{M}  = \sum_{\gamma \in H_1(M)} {  \int D \omega \, e^{i S[\widehat A_\gamma + \omega]} \, W_L ( \widehat  A_\gamma + \omega ) \over  \int D \omega \, e^{i S[ \omega]} }  \; ,
\label{5.6}
\ee
let us choose the background origins $\widehat A_\gamma$. Each element $\gamma \in H_1(M)$ can be decomposed as
\be
\gamma = \gamma_\varphi +  \gamma_\tau  \; ,
\label{5.7}
\ee
where $\gamma_\varphi \in F(M)$ and $\gamma_\tau \in T(M)$. In particular, one can write
\be
\gamma_\varphi = z_1 g_1 + z_2 g_2 + \cdots + z_B g_B \quad  , \quad  \gamma_\tau = n_1 h_1 + n_2 h_2 + \cdots + n_N h_N \; ,
\label{5.8}
\ee
for integers $z_i \in {\mathbb Z}$ and $n_j$, with $0 \leq n_j \leq p_j -1$. Accordingly one can put
\be
\widehat A_\gamma = \widehat A_{\gamma_\varphi} + \widehat A_{  \gamma_\tau} \; ,
\label{5.9}
\ee
where
\be
\widehat A_{\gamma_\varphi} = z_1 \eta_{1} + z_2 \eta_{2} + \cdots + z_B \eta_B \; ,
\label{5.10}
\ee
and
\be
 \widehat A_{  \gamma_\tau} = n_1 A^0_1 + n_2 A^0_2 + \cdots + n_N A^0_N \; .
 \label{5.11}
\ee
The torsion components $ \widehat A_{  \gamma_\tau} $  represent the canonical origins which describe the gauge orbit associated with the flat connections of type (\ref{4.4}).  In particular, the class $A^0_j$ (with $j=1,2,..., N$) denotes the gauge orbit corresponding to the generator $h_j$ of $T(M)$,
\be
A^0_j \leftrightarrow \{ 0 , \Lambda_{ab}( h_j ) , N_{abc} (h_j) \} \; .
\label{5.12}
\ee
The fibres of $H^1_D(M)$ over $H_1(M)$ which are labelled by the elements $\gamma_\varphi \in F(M)$ do not possess a canonical origin and, in order to simplify the exposition, the choice of $\widehat A_{\gamma_\varphi} $ illustrated in equation (\ref{5.10}) is based on the distributional DB classes   $\eta_i $ (with $i=1,2,.., B$)  which can be represented by
\be
\eta_i \leftrightarrow \{ \alpha_a (C_{g_i} ), \Lambda_{ab} (C_{g_i}), N_{abc} (C_{g_i}) \} \; ,
\label{5.13}
\ee
where $ \alpha_a (C_{g_i})$ is a de~Rham-Federer 1-current defined in ${\cal U}_a$ such that $ d\alpha_a (C_{g_i})$ has support on the  restriction  in the open ${\cal U}_a$ of a knot $C_{g_i}\subset M$ that represents the generator $g_i$ of $F(M)$. It is convenient to introduce a framing for each knot $C_{g_i}$, so that all expressions containing the distributional DB class $\widehat A_{\gamma_\varphi}$ are well defined.  The final expression that will be obtained for $\langle\!  \langle  W_L(A) \rangle \! \rangle \big |_{M} $ does not  depend on the choice of the framing of $C_{g_i}$.

\subsection {Zero modes integration}
Each gauge orbit is then denoted by
\be
\widehat A_\gamma + \omega  = \widehat A_{\gamma_\varphi} +  \widehat A_{\gamma_\tau } + \omega^0 + \widetilde \omega  \; ,
\label{5.14}
\ee
and the functional integration takes the form
\bea
\sum_{\gamma \in H_1(M)} && {\hskip -0.7 truecm} \int D \omega \; F[\widehat A_\gamma + \omega] = \nonumber \\
  && {\hskip -1.9 truecm}= \! \! \sum_{\gamma_\tau \in T(M)}  \sum_{z_1= - \infty }^{+ \infty }  \! \! \cdots  \! \! \sum_{z_B = - \infty }^{+ \infty } \int_0^1 \! \! d \theta_1 \cdots \int_0^1 \! \! d \theta_B \! \int D\widetilde \omega \; F[ \widehat A_{\gamma_\varphi } +  \widehat A_{\gamma_\tau } + \omega^0 + \widetilde \omega \, ] \, .
\label{5.15}
\eea
 We now need to determine the dependence of the action $S[\widehat A_\gamma + \omega] $ and of the holonomy $W_L[\widehat A_\gamma + \omega]$ on the field components (\ref{5.14}). One has
\bea
S[\widehat A_\gamma + \omega] &=& S[\widehat A_{\gamma_\varphi } +  \widehat A_{\gamma_\tau} + \omega^0 + \widetilde \omega]= \nonumber \\
&=& S[\widehat A_{\gamma_\tau}  + \widetilde \omega] + 4 \pi k \int_M \left [ (\widehat A_{\gamma_\tau}  + \widetilde \omega) * (\widehat A_{\gamma_\varphi } + \omega^0 ) \right ] \nonumber \\
&& \quad + 2 \pi k \int_M \left [ \widehat A_{\gamma_\varphi } * \widehat A_{\gamma_\varphi } + \omega^0 * \omega^0 + 2 \, \omega^0 * \widehat A_{\gamma_\varphi} \right ] \; .
\label{5.16}
\eea
Since the first component of the representative connections (\ref{5.12}) is vanishing, whereas only the first component of the representative connections (\ref{4.9}) is not vanishing, one gets
\be
\int_M \widehat A_{\gamma_\tau } * \omega^0 = 0 \qquad \hbox{mod }{\mathbb Z}  \; .
\label{5.17}
\ee
For the reason that $\widetilde \omega \in
 \Omega^1(M) / \Omega^1_{\mathbb Z} (M)$, $\omega^0 \in \Omega^1(M) / \Omega^1_{\mathbb Z} (M)$ and $d\omega^0 =0$, one finds
\be
\int_M \widetilde \omega * \omega^0  = \int_M  \omega^0 * \omega^0 = 0  \qquad \hbox{mod } {\mathbb Z} \; .
\label{5.18}
\ee
The framing procedure, which defines the self-linking numbers, produces integer values for  the  self-interactions of  the distributional DB  classes $A_{\gamma_\varphi }$, thus
\be
\int_M \widehat A_{\gamma_\varphi } * \widehat A_{\gamma_\varphi } = 0 \qquad \hbox{mod }   {\mathbb Z}   \; .
\label{5.19}
\ee
The normalization condition (\ref{4.6}) and the definitions (\ref{4.9}) and (\ref{5.10}) imply
\be
\int_M \omega^0 * \widehat A_{\gamma_\varphi } = \sum_{i=1}^{N_F} z_i \theta_i   \qquad \hbox{mod } {\mathbb Z} \; .
\label{5.20}
\ee
Therefore
\be
\exp \left ( i S[\widehat A_\gamma + \omega] \right ) = \exp \left ( \! i S[\widehat A_{\gamma_\tau }  +  \widetilde \omega] + 4 i \pi k \! \! \int_M \! (\widehat A_{\gamma_\tau }  + \widetilde \omega) * \widehat A_{\gamma_\varphi} + 4 i \pi k \sum_i z_i \theta_i \! \right ) \, .
\label{5.21}
\ee
Let us now consider the holonomy
\be
W_L[\widehat A_\gamma + \omega ] = \exp \! \left ( \! 2 \pi i \int_M [\widehat A_\gamma + \omega ] * \eta_L \! \right ) = \exp \! \left ( 2 \pi i \!  \int_M \! \left [ \widehat A_{\gamma_\varphi } +  \widehat A_{\gamma_\tau } + \omega^0 + \widetilde \omega \right ] * \eta_L  \! \right ) \, .
\ee
The distributional DB classes have integer linking
\be
\int_M \widehat A_{\gamma_\varphi } * \eta_L =
0 \qquad \hbox{mod } {\mathbb Z}  \; ,
\label{5.22}
\ee
and condition (\ref{4.6}) together with the definition of the homology classes  (\ref{5.4}) and (\ref{5.5}) give
\be
\int_M \omega^0 * \eta_L = \sum_{i=1}^B a^i_L \theta_i \qquad \hbox{mod } {\mathbb Z} \; .
\label{5.23}
\ee
Consequently
\be
\exp \left ( 2 \pi i \int_M  (\widehat A_\gamma + \omega ) * \eta_L \right ) = e^{2 \pi i \sum_i a^i_L \theta_i } \, \exp \left ( 2 \pi i \int_M  (\widehat A_{\gamma_\tau } + \widetilde \omega ) * \eta_L \right ) \; .
\label{5.24}
\ee
The expectation value (\ref{5.6}) then becomes
\bea
 \langle\!  \langle  W_L(A) \rangle \! \rangle \Big |_{M}  \! \! &=&\! \!  \sum_{\gamma_\tau \in T(M)}  \sum_{z_1= - \infty }^{+ \infty }\! \! \cdots  \! \! \sum_{z_B= - \infty }^{+ \infty } \int_0^1 \! \! d \theta_1 \cdots \int_0^1 \! \! d \theta_B \,
 e^{2 \pi i \sum_j [ 2 k z_j + a^j_L ] \theta_j } \times \nonumber \\
 && \quad \times  {\int  D\widetilde \omega \, e^{iS[\widehat A_{\gamma_\tau}+ \widetilde \omega ] } e^{2 \pi i \int (\widehat A_{\gamma_\tau}+ \widetilde \omega) * \eta_L} e^{4 \pi i k \int (\widehat A_{\gamma_\tau}+ \widetilde \omega) * \widehat A_{\gamma_\varphi } } \over \int D \omega \, e^{i S[\omega ]}}\;  .
\label{5.25}
\eea
Each single integral in the $\theta_j$ variable gives
\be
\int_0^1 d \theta_j \> e^{2 \pi i  [ 2 k z_j + a^j_L ] \theta_j } = \delta (2 k z_j + a^j_L )\; .
\label{5.26}
\ee
Both $z_j $ and $a_L^j$ are integers, and the  constraint (\ref{5.26}) is satisfied provided
$a^j_L \equiv 0 $~mod~$2k$. Thus, in order to have $\langle\!  \langle  W_L(A) \rangle \! \rangle \big |_{M} \not= 0 $, it must be  $
[L]_F \equiv 0 $~mod~$2k$, that is
\be
a_L^j \equiv 0 \qquad \hbox{mod }2k \quad , \quad \forall j = 1,2,..., B \; .
\label{5.27}
\ee
When $ [L]_F \equiv 0 $~mod~$2k$, in expression (\ref{5.25}) the sums over the $z$-variables have the effect of replacing each variable $z_j$ by $\overline z_j$ given by
\be
z_j \rightarrow \overline z_j = - ( a_L^j/2k ) \; .
\label{5.28}
\ee
From the definition (\ref{5.10}) it follows then
\be
\widehat A_{\gamma_\varphi } \Big |_{z_j = \overline z_j} =  {1\over 2k } \, \eta_{ L_\bullet } \quad , \quad \eta_{ L_\bullet }=  - a_L^1 \eta_1 - a_L^2 \eta_2 - \cdots - a_L^B \eta_B  \; ,
\label{5.29}
\ee
where $\eta_{ L_\bullet }$ can be interpreted as  the distributional DB class which is associated with the oriented framed colored link $L_\bullet = C_{g_1} \cup C_{g_2}\cup \cdots \cup C_{g_{N_F}} \subset M$ in which the component $C_{g_j}$ has color given by the integer charge $- a_L^j$.  So from equation (\ref{5.25}) one obtains
\be
 \langle\!  \langle  W_L(A) \rangle \! \rangle \Big |_{M}  = \sum_{\gamma_\tau \in T(M)}     {\int  D\widetilde \omega \, e^{iS[\widehat A_{\gamma_\tau }+ \widetilde \omega ] } e^{2 \pi i \int (\widehat A_{\gamma_\tau }+ \widetilde \omega) * ( \eta_L + \eta_{L_\bullet } )  } \over \int D \omega \, e^{i S[\omega ]}}    \;  .
\label{5.30}
\ee
 The distributional DB class
 \be
 \eta_{L_\tau } \equiv  \eta_L + \eta_{ L_\bullet }
 \label{5.31}
 \ee
  is  associated with the link
\be
L_\tau = L \cup { L_\bullet } \subset M \; ,
\label{5.32}
\ee
and the homology class $[L_\tau ]$ of $L_\tau $ has nontrivial components in the torsion subgroup exclusively, more precisely
\be
[L_\tau ] = [L]_T  = \sum_{i=1}^N b_L^i \, h_i \; .
\label{5.33}
\ee

\begin{rmk}
The generators of the torsion subgroup $T(M)$ are not linked with the generators of $F(M)$, therefore in the computation of expression (\ref{5.30}) the components of $ L_\bullet $ supply  various integer linking numbers between $C_{g_i}$ and $C_{g_j}$ (for arbitrary $i$ and $j$) and  between $C_{g_i}$ and the $L$ components. In particular, the contribution of $ L_\bullet $ to the integral (\ref{5.30}), which depends on the framing of the knots $C_{g_j} $ exclusively, is given by the multiplicative factor  $\exp [ -(2 \pi i / 4k)\sum_j (a_L^j)^2 \ell k (C_{g_j}, C_{g_j {\rm f}}) ]$ which is of the type (\ref{2.9}).  Consequently, since each $a_L^j$ is a multiple of $2k$, $\langle\!  \langle  W_L(A) \rangle \! \rangle \big |_{M}$ does not depend on the  choice of the framing of the knots $\{ C_{g_j} \}$.
\end{rmk}

\begin{rmk}
Since  the homology class of $L_\tau $ has no components in the group $F(M)$, instead of integrating over $\widetilde \omega $, in the functional integral (\ref{5.30}) one can integrate directly   over the whole space of the variables $\omega \in \Omega^1(M)/ \Omega^1(M)_{\mathbb Z}$  without modifying the result; indeed the integral over the amplitudes of the zero modes  simply gives a multiplicative unit factor. This is a consequence of the fact that each  amplitude $\theta_j $ of the zero modes takes values in the  range $0 < \theta_j \leq 1$.
\end{rmk}

\noindent Thus the outcome (\ref{5.30}) can also be written in the following way:
\be
\langle\!  \langle  W_L(A) \rangle \! \rangle \Big |_{M}  = 0 \quad , \quad \hbox{ if } \; [L]_F
\not\equiv 0 \; \; \hbox{ mod } 2k \; ,
\label{5.34}
\ee
and when $ [L]_F \equiv 0 $~mod~$2k$ one gets
\be
 \langle\!  \langle  W_L(A) \rangle \! \rangle \Big |_{M}  = \sum_{\gamma_\tau \in T(M)}     {\int  D  \omega \, e^{iS[\widehat A_{\gamma_\tau} +  \omega ] } e^{2 \pi i \int (\widehat A_{\gamma_\tau}+  \omega) *  \eta_{L_\tau  } }\over \int D \omega \, e^{i S[\omega ]}}    \;  .
\label{5.35}
\ee
In view of equations (\ref{5.32}) and (\ref{5.33}), one can summarize the results (\ref{5.34}) and (\ref{5.35}) by saying that the functional integration over the zero-modes flat connections acts as a projection into the sector of vanishing free homology.

\subsection {Factorization}
The action $S[\widehat A_{\gamma_\tau} +  \omega]$ is given by
\be
S[\widehat A_{\gamma_\tau} +  \omega] = S[\widehat A_{\gamma_\tau} ] + S[ \omega] + 4 \pi k \int_M \omega * \widehat A_{\gamma_\tau}  \; ;
\label{5.36}
\ee
since $\omega \in \Omega^1(M)/ \Omega^1(M)_{\mathbb Z}$ and the canonical origins $A^0_j $ are represented by the connections (\ref{5.12}), it follows
\be
\int_M \omega * \widehat A_{\gamma_\tau}  = 0 \qquad \hbox{mod }{\mathbb Z}  \; .
\label{5.37}
\ee
Therefore equation (\ref{5.35}) takes the form
\be
 \langle\!  \langle  W_L(A) \rangle \! \rangle \Big |_{M}  = \left ( \sum_{\gamma_\tau \in T(M)} e^{iS[\widehat A_{\gamma_\tau}  ] }  e^{2 \pi i \int \widehat A_{\gamma_\tau} *  \eta_{L_\tau  } }    \right ) {\int  D  \omega \, e^{iS[  \omega ] } e^{2 \pi i \int   \omega *  \eta_{L_\tau  } }\over \int D \omega \, e^{i S[\omega ]}}    \;  .
\label{5.38}
\ee
This expression  shows that, as a consequence of equation (\ref{5.37}),  the path-integral over $\omega $ and the sum over the torsion background fields given by the canonical origins $A^0_j$ factorize.
The term
\be
{\int  D  \omega \, e^{iS[  \omega ] } e^{2 \pi i \int   \omega *  \eta_{L_\tau  } }\over \int D \omega \, e^{i S[\omega ]}}  = e^{- (2 \pi i /4 k)\,  \Lambda_M (L_\tau , L_\tau) }
\label{5.39}
\ee
is called the perturbative component of $ \langle\!  \langle  W_L(A) \rangle \! \rangle \big |_{M} $ because it coincides with its Taylor expansion in powers  of the variable $1 / k$ and it assumes the unitary value in the  $1 / k \rightarrow 0$ limit.  The integral (\ref{5.39}) is the analogue of expression (\ref{2.9}); the quadratic function  $\Lambda_M (L_\tau , L_\tau)$  of the link $L_\tau$  assumes rational values and  can be defined  in terms of appropriate linking numbers. On the other hand, the term
\be
\sum_{\gamma_\tau \in T(M)} e^{iS[\widehat A_{\gamma_\tau}  ] }  e^{2 \pi i \int \widehat A_{\gamma_\tau} *  \eta_{L_\tau  } }   = \sum_{n_1 =0}^{p_1 -1} \cdots \sum_{n_N =0 }^{p_N -1}  e^{2 \pi i k \sum_{ij} n_i n_j \int A^0_i  *   A^0_j }  e^{2 \pi i \sum_j n_j \int  A^0_j *  \eta_{L_\tau  } }
\label{5.40}
\ee
does not admit a power expansion in powers of $1/k$ around $1/k =0 $ and it represents the nonperturbative component  of $ \langle\!  \langle  W_L(A) \rangle \! \rangle \big |_{M} $.   So the gauge orbits  $\widehat A_{\gamma_\tau}$ of the torsion flat connections regulate the non-perturbative contributions to the expectation values.

\subsection {Perturbative component}  The path-integral (\ref{5.39}) can be computed by using a procedure which is  similar to the method illustrated in \S~2.1 and \S~2.2. In order to simplify the exposition, it is convenient  to use two properties of the CS path-integral  according to which one can  replace  the  link $L_\tau$ by an appropriate single oriented framed  knot $K_L$ with color specified by the unitary charge $q=1$.

\begin{enumerate}[(a)]

\item The first property \cite{9} reads
\be
{\int  D  \omega \, e^{iS[  \omega ] } e^{2 \pi i \int   \omega *  \eta_{L_\tau  } }\over \int D \omega \, e^{i S[\omega ]}}  = {\int  D  \omega \, e^{iS[  \omega ] } e^{2 \pi i \int   \omega *  \eta_{ L^*_\tau  } }\over \int D \omega \, e^{i S[\omega ]}} \; ,
\label{5.41}
\ee
where  $ L^*_\tau \subset M $ is the simplicial satellite  of $L_\tau$. The oriented framed colored link $ L^*_\tau  $ is obtained from $L_\tau $ by replacing each component $K_j$ of $L_\tau $, that has color given by the  charge $q_j\not= \pm 1$, by  $| q_j |$ parallel copies of $K_j$ with unit charge; these parallel copies of $K_j$ ---together with their framings--- belong to the band which is bounded by $K_j$ and by its framing $K_{j { \rm f}}$. Thus, with a suitable choice for the orientations  of the link components, all the components of $ L^*_\tau $ have unit charge $q=1$.  Property (\ref{5.41}) follows from  the definition of the framing procedure \cite{9,18}.

\item The second property \cite{1} states that
\be
{\int  D  \omega \, e^{iS[  \omega ] } e^{2 \pi i \int   \omega *  \eta_{L^*_\tau  } }\over \int D \omega \, e^{i S[\omega ]}}  = {\int  D  \omega \, e^{iS[  \omega ] } e^{2 \pi i \int   \omega *  \eta_{  K_{L} } }\over \int D \omega \, e^{i S[\omega ]}} \; ,
\label{5.42}
\ee
where the oriented framed  knot $K_L \subset M $ (with color $q=1$) is the band connected sum \cite{1,19} of all the components of $L^*_\tau $. The sum of two knots is illustrated in Figure~5.1. Property (\ref{5.42}) is a consequence of the fact  if one adds or eliminates one unknot ---which belongs to a 3-ball in $M$ and has trivial framing--- the  expectation values of the link holonomies are left invariant.
\end{enumerate}

 By construction, the homology class $[K_L]$ of the knot $K_L$ is equal to the homology class of the link $L_\tau$. Let us now consider the following  two possibilities.

\vskip 0.8 truecm
\centerline {\includegraphics[width=3.80 in]{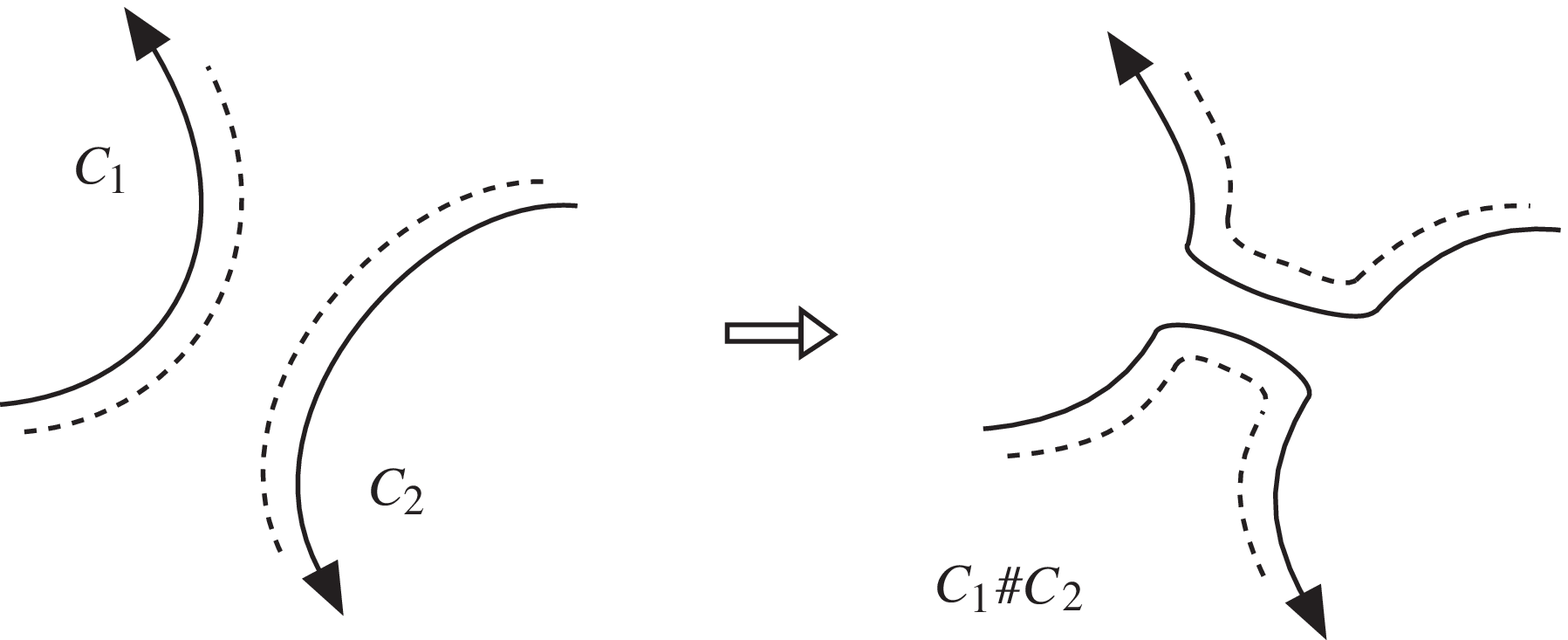}}
\vskip 0.3 truecm
\centerline {{Figure 5.1.} {Band connected sum $C_1 \# C_2$ of the knots $C_1$ and $C_2$.}}

\vskip 0.6 truecm

\subsubsection {Trivial homology} If $[L_\tau ] = [K_L] =0$, one can find a Seifert surface $\Sigma \subset M $ for the knot $K_L$ and  define the associated 1-current $\alpha_\Sigma$ such that $\int_M \omega * \eta_{K_L} = \int_M \omega \wedge d \alpha_\Sigma $.
Note that the current  $\alpha_\Sigma$ is globally  defined in the manifold $M$, so the product $ (1/ 2k ) \, \alpha_\Sigma  $ is  well defined. Then in the path-integral (\ref{5.42}) one can perform the change of variables
\be
\omega = \eta^\prime_{K_L} + \omega^\prime \; ,
\label{5.43}
\ee
where the class  $\eta^\prime_{K_L} $ is represented by
\be
 \eta^\prime_{K_L} \leftrightarrow \left \{  -   \left ( {\alpha_\Sigma \over 2k }\right )_a , 0 , 0 \right \} \; ,
 \label{5.44}
 \ee
 and $\omega^\prime $ designates the fluctuating variable. The restriction of $ (\alpha_\Sigma / 2k )$ in the open domain ${\cal U}_a$ has been denoted by $( \alpha_\Sigma / 2k )_a$. Since $e^{iS[\omega] } e^{2 \pi i \int \omega * \eta_{K_L}}  = e^{i S[\omega^\prime]} e^{- 2 \pi i / 4k \int \eta^\prime_{K_L} * \eta^\prime_{K_L} }$, in expression (\ref{5.42})  the functional integration  over $\omega^\prime $
 factorizes in the numerator and cancels with the denominator. So, by taking into account equations (\ref{5.41}) and (\ref{5.42}), one obtains
 \be
{\int  D  \omega \, e^{iS[  \omega ] } e^{2 \pi i \int   \omega *  \eta_{L_\tau  } }\over \int D \omega \, e^{i S[\omega ]}}  = e^{- (2 \pi i /4 k)\,  \ell k (K_L , K_{L {\rm f} } ) }    \; ,
\label{5.45}
\ee
 where the linking number $\ell k (K_L , K_{L {\rm f}})$ ---which takes integer values--- is well defined because $[K_L] = [K_{L {\rm f}}] = 0$.

 \subsubsection {Nontrivial torsion} When $[L_\tau ] = [K_L] \in T(M)$ (with $[K_L]  \not= 0$), on can always find a integer $p\in {\mathbb Z} $ such that $p   [K_L] =0  $.  So, let us consider the satellite of $K_L$ which is made of $p$ parallel copies of the knot $K_L$ (each copy belongs to the band bounded by $K_L$ and its framing $K_{L {\rm f}}$), the band connected sum of all these parallel knots defines a framed oriented knot $K^p_L \subset M$ with $[K^p_L] =0$.  We call $K^p_L $ the $p$-covering of the knot $K_L$.
 Let $\Sigma^\prime \subset M$ be a Seifert surface of $K^p_L $ and let $\alpha_{\Sigma^\prime}$ be the corresponding 1-current. Again,   the current  $\alpha_{\Sigma^\prime }$ is globally  defined in the manifold $M$, so the product $(1 / p) \alpha_\Sigma$ is well defined.
 Let us introduce the distributional class $\eta_{K^p_L}$ which satisfies
\be
\eta_{K^p_L} \leftrightarrow \left \{ {1\over p} \left ( \alpha_{\Sigma^\prime} \right )_a , 0 , 0 \right \} \; .
\label{5.46}
\ee
Then
\be
{\int  D  \omega \, e^{iS[  \omega ] } e^{2 \pi i \int   \omega *  \eta_{K_L  } }\over \int D \omega \, e^{i S[\omega ]}}  = {\int  D  \omega \, e^{iS[  \omega ] } e^{2 \pi i \int   \omega *  \eta_{K^p_L} }\over \int D \omega \, e^{i S[\omega ]}} \; ,
\label{5.47}
\ee
and from now on one can proceed  as in the trivial homology case.  Consequently one finds
\be
{\int  D  \omega \, e^{iS[  \omega ] } e^{2 \pi i \int   \omega *  \eta_{L_\tau  } }\over \int D \omega \, e^{i S[\omega ]}}  = e^{- (2 \pi i /4 k)\,   \ell k (K^p_L , K^p_{L {\rm f} } )/p^2 }    \; .
\label{5.48}
\ee

\subsection {Nonperturbative component} For each canonical origin $\widehat A_{\gamma_\tau }$ (with $\gamma_\tau \in T(M)$),  the amplitude
\be
 e^{iS[\widehat A_{\gamma_\tau}  ] }   =  e^{2 \pi i k \sum_{ij} n_i n_j \int A^0_i  *   A^0_j }   = e^{2 \pi i k \sum_{ij} n_i n_j Q_{ij}}
\label{5.49}
\ee
determines a  ${\mathbb Q} / {\mathbb Z}$-valued quadratic form $Q$ on $T(M)$ which is specific of  the  manifold $M$.
The value of the CS action  $S[\widehat A_{\gamma_\tau } ]$  can be computed by using different methods \cite{1,22,23,24}; in particular, $S[\widehat A_{\gamma_\tau } ]$ can also be interpreted  as an appropriate linking number. For each element $\gamma_\tau $ of the torsion group
one can choose a representative oriented  knot $C_{\gamma_\tau} \subset M$. Let $C_{\gamma_\tau {\rm f}} $ be a framing for $C_{\gamma_\tau}$. The  self-linking number of $C_{\gamma_\tau}$ ---which is equal to the linking number of $C_{\gamma_\tau}$ with $C_{\gamma_\tau {\rm f}} $---  modulo integers determines the value of $Q( \gamma_\tau )$. This linking number can be computed by using the method illustrated in \S~5.4. Namely, if  $p \, \gamma_\tau = 0$ for a given integer  $p \in {\mathbb Z}$,  consider the framed satellite of $C_{\gamma_\tau}$ made of $p$ parallel copies of  the framed knot $C_{\gamma_\tau}$  that belong to the band which is bounded by $C_{\gamma_\tau}$ and $C_{\gamma_\tau {\rm f}} $; finally the  sum of all these components defines a framed knot $C^p_{\gamma_\tau}$. Since $C^p_{\gamma_\tau}$ has trivial homology, $[C^p_{\gamma_\tau}] = 0$, there exists a Seifert surface $\Sigma \subset M$ of $C^p_{\gamma_\tau}$ and one can define the corresponding  de~Rham-Federer 1-current $\alpha_\Sigma$.   Let $\alpha_{\Sigma_{\rm f}}$ be the 1-current which is associated with a Seifert surface $\Sigma_{\rm f}$ of the framing of $C^p_{\gamma_\tau}$. Then the self-linking number of $C_{\gamma_\tau}$ is given by
\be
\ell k (C_{\gamma_\tau}, C_{\gamma_\tau {\rm f}}) = {1\over p^2} \int_M \alpha_\Sigma \wedge d \alpha_{\Sigma_{\rm f}} = {1\over p^2} \int_M \alpha_{\Sigma_{\rm f}} \wedge d \alpha_\Sigma \; ,
\label{5.50}
\ee
and assumes rational values in general. One has
\be
 e^{iS[\widehat A_{\gamma_\tau}  ] }   =  e^{- 2 \pi i k \, \ell k (C_{\gamma_\tau}, C_{\gamma_\tau {\rm f}})}   = e^{- (2 \pi i k / p^2 )  \int_M \alpha_\Sigma \wedge d \alpha_{\Sigma_{\rm f}} } =  e^{2 \pi i k \, Q(\gamma_\tau )}  \; .
\label{5.51}
\ee
Given a integer Dehn surgery presentation of $M$, the quadratic form $Q$ can also be derived \cite{1,6,25} from the expression of the linking matrix of the surgery instructions.

\begin{rmk}  Since the CS coupling constant $k$ takes integer values, the quadratic form $Q(\gamma_\tau )$ ---which is determined by equation (\ref{5.51}) for arbitrary integer $k$---  is defined modulo integers. Moreover the value of the amplitude $e^{iS[\widehat A_{\gamma_\tau}  ] }$ does not depend on the particular choice of the framing $C_{\gamma_\tau {\rm f}} $. Indeed, under a modification of the framing $C_{\gamma_\tau {\rm f}} $, the variation of the intersection number $\int_M \alpha_\Sigma \wedge d \alpha_{\Sigma_{\rm f}}$ is given by
\be
\Delta \left ( \int_M \alpha_\Sigma \wedge d \alpha_{\Sigma_{\rm f}}  \right ) = p^2 \times \hbox{ integer } \; ,
\label{5.52}
\ee
because  the knot $C^p_{\gamma_\tau}$ is the band connected sum of $p$ parallel copies of $C_{\gamma_\tau } $.  The change (\ref{5.52}) of the self-linking number $\int_M \alpha_\Sigma \wedge d \alpha_{\Sigma_{\rm f}}$ leaves expression (\ref{5.51}) invariant.
\end{rmk}

\noindent Finally the value of the amplitude
\be
 e^{2 \pi i \int \widehat A_{\gamma_\tau} *  \eta_{L_\tau  } }   =  e^{2 \pi i \sum_j n_j \int  A^0_j *  \eta_{L_\tau  } }
\label{5.53}
\ee
can be determined by computing the linking numbers of  the components of the link $L_\tau $ with the representative knots of the generators of the torsion group. In this calculation also one can use the methods illustrated above; the various linking numbers generally assume rational values.

\subsection {Path-integral invariants} The result of the functional integration can be summarized as
\bea
 \langle\!  \langle  W_L(A) \rangle \! \rangle \Big |_{M}  &=&  \delta ([L]_F \equiv 0 \hbox{ ~mod } 2k ) \times e^{- (2 \pi i /4 k)\,   \ell k (K^p_L , K^p_{L {\rm f} } )/p^2 }   \times \nonumber \\
   && \times
 \left ( \sum_{n_1 =0}^{p_1 -1} \cdots \sum_{n_N =0 }^{p_N -1}  e^{2 \pi i k \sum_{ij} n_i n_j Q_{ij}}  e^{2 \pi i \sum_j n_j \int  A^0_j *  \eta_{L_\tau  } }
  \right )    \;  ,
\label{5.54}
\eea
where all the various functions which appear in the exponents represent appropriate linking numbers.  By inserting $L=0$ in equation (\ref{5.54}), one obtains  the path-integral partition function
\be
 Z_k (M) \equiv \langle\!  \langle  1\rangle \! \rangle \Big |_{M}  =    \sum_{n_1 =0}^{p_1 -1} \cdots \sum_{n_N =0 }^{p_N -1}  e^{2 \pi i k \sum_{ij} n_i n_j Q_{ij}}     \;  .
\label{5.55}
\ee

\section {Comparison with the Reshetikhin-Turaev surgery invariants}

 Let us briefly recall the definition of the abelian surgery invariants of Reshetikhin-Turaev \cite{5,6,7,8}. Each closed oriented 3-manifold admits a integer Dehn surgery presentation in $S^3$, in which the surgery instruction  is described  by a framed link in $S^3$.  Suppose that the framed surgery link ${\cal L} \subset S^3$, which corresponds to the 3-manifold $M_{\cal L}$, has components ${\cal L} =  {\cal L}_1 \cup {\cal L}_2 \cup \cdots  \cup {\cal L}_{m}$. With the introduction of a orientation for each component of $\cal L$, one  can define
 the surgery function $\widehat W_{\cal L} (A)$  by means of the equation
\be
\widehat W_{\cal L} (A) =  \sum_{q_1 =0}^{2 k -1} e^{2 \pi i q_1 \oint_{{\cal L}_1}\! \! A } \, \sum_{q_2 =0}^{2 k -1}e^{2 \pi i q_2 \oint_{{\cal L}_2} \! \! A } \cdots \sum_{q_m =0}^{2 k -1}e^{2 \pi i q_m \oint_{{\cal L}_m} \! \!  A } \; ,
\label{6.1}
\ee
where $\exp \left ( 2 \pi i q_j \oint_{{\cal L}_j  }  A \right )$ denotes the gauge holonomy associated with the component ${\cal L}_j$ with charge $q_j$.
Let $ \widetilde  {\mathbb L}$ be the linking matrix of the surgery link, and let  $\sigma ({\cal L}) $ denote the signature of $ \widetilde  {\mathbb L}$. For fixed integer $k$, the following  combination $I_k(M_{\cal L} )$ of expectation values of the sphere,
\bea
I_k(M_{\cal L} ) &=& \left ( 2 k \right )^{-m/2} e ^{i \pi \sigma ({\cal L}) /4}\, \langle \, \widehat W_{\cal L} (A) \rangle \Big |_{S^3}\nonumber \\
&=& \left ( 2 k \right )^{-m/2} e ^{i \pi \sigma ({\cal L}) /4}\, \sum_{q_1 =0}^{2 k -1} \cdots  \sum_{q_m =0}^{2 k -1} e^{- (2 \pi i / 4 k ) \sum_{ij=1}^m q_i q_j \widetilde { \mathbb L}_{ij}} \; ,
\label{6.2}
\eea
is invariant under Kirby moves \cite{14,19} and thus it represents a topological invariant of the oriented manifold $ M_{\cal L}$.    Similarly, if $L$ denotes a framed oriented colored link in the complement of ${\cal L}$ in  $S^3$, then
\be
I_k(M_{\cal L} ; L) = \left ( 2 k \right )^{-m/2} e ^{i \pi \sigma ({\cal L}) /4}\, \langle  \, \widehat W_{\cal L} (A) \, W_L(A)\rangle \Big |_{S^3}
\label{6.3}
\ee
defines a surgery invariant of the link $L$ in the 3-manifold $M_{\cal L}$. The defining expressions (\ref{6.2}) and (\ref{6.3}) are not the result of a path-integral computation  in the 3-manifold $M_{\cal L}$. The abelian Reshetikin-Turaev invariants (\ref{6.2}) and (\ref{6.3})  are defined by means  of  appropriate combinations  of the  link invariants  of the sphere $S^3$ in which one of the links is the surgery link; for this reason expressions (\ref{6.2}) and (\ref{6.3}) are called surgery invariants.

Expressions (\ref{6.2}) and (\ref{6.3}) can be transformed by means of the Deloup-Turaev reciprocity formula \cite{25}.
The symmetric bilinear form on the lattice $W$ of Theorem~1  contained in Ref.\cite{25}  corresponds to the bilinear form which is defined by the linking matrix $\widetilde {\mathbb L}$, and the sum over the elements in the  dual lattice $W^{\bullet}$ is in agreement with the sum over the elements of the torsion group $T(M)$.  The vanishing eigenvalues of $\widetilde {\mathbb L}$ are correlated  with the gauge orbits (\ref{4.9}) of flat connections $\omega^0 (\theta) \in   \Omega^1(M) / \Omega^1_{\mathbb Z} (M)$ due to the zero modes. Whereas the quadratic form $Q$ can be understood \cite{1} as a suitable  inverse of the minor of $\widetilde {\mathbb L}$  in the torsion subspace.  As a consequence of  Theorem~1 of Ref.\cite{25},  one has
\be
 \langle\!  \langle  W_L(A) \rangle \! \rangle \Big |_{M_{\cal L}}  = \left ( 2k \right )^{-B/2} \, (p_1 p_2 \cdots p_N)^{1/2} \, I_k(M_{\cal L} ; L) \; .
 \label{6.4}
\ee
In particular, as far as the partition function is concerned, equation (\ref{6.4}) gives
\be
Z_k(M_{\cal L})  = \left ( 2k \right )^{-B/2} \, (p_1 p_2 \cdots p_N)^{1/2} \, I_k(M_{\cal L} ) \; .
 \label{6.5}
\ee
So, when $Z_k(M_{\cal L}) \not= 0$, the standardly normalized path-integral expectation values (\ref{3.8}) coincide \cite{26} with the ratios of the Reshetikhin-Turaev invariants
\be
  \langle  W_L(A)  \rangle \Big |_{M_{\cal L}}  =  {  \langle\!  \langle  W_L(A) \rangle \! \rangle \Big |_{M_{\cal L}}  \over  Z_k(M_{\cal L}) }  = {I_k(M_{\cal L} ; L)  \over I_k(M_{\cal L} ) }  \; .
\label{6.6}
\ee

\section {Examples}

The effects of the topology of the manifold $M$ on the path-integral invariants (\ref{5.54}) are of two types. The freely component $F(M)$ of the homology group is simply related with the
$\delta ([L]_F \equiv 0 \hbox{ ~mod } 2k )$ factor.  Whereas the nontrivial topology  contribution is described by the quadratic form $Q$ on the torsion group $T(M)$. So let us present examples of 3-manifolds with homology groups containing the torsion component  exclusively.

Let us consider the  lens spaces $L_{p/r} $ in which the two coprime integers $p$ and $r$ satisfy $ p > 1 $ and  $1 \leq r < p$. When $p \not= p^\prime$, the lens spaces $L_{p/r}$ and $L_{p^\prime /r^\prime }$ are not homeomorphic;   the manifolds $L_{p/r}$ and $L_{p/r^\prime }$  are homeomorphic iff $\pm r^\prime \equiv r^{\pm 1} $ (mod~$p$).      The fundamental group is abelian $\pi_1(L_{p/r}) = {\mathbb Z}_p  $ and coincides with the homology group $H_1 (L_{p/r}) = T (L_{p/r}) = {\mathbb Z}_p  $. A generic element $\gamma \in T (L_{p/r})$ can be written as $\gamma = n h$ where $h$ is the generator of $T (L_{p/r})$ and  satisfies  $p h =0$.
  The manifold $L_{p/r}$ admits \cite{27} a surgery presentation in $S^3$ in which the surgery instruction is given by the unknot $U\subset S^3$ with surgery coefficient $p/r$.  Let $V$ be a tubular neighbourhood of $U$; $L_{p/r}$ is obtained by removing the interior $\vnt $ of $V$ from $S^3$ and by gluing $V$ with $S^3 - \vnt $ according to a  homemorphism $ f^* : \der V \rightarrow \der (S^3 - \vnt )$ which sends a meridian $\mu$ of $V$ into a $(p,r)$ torus knot in $\der (S^3 - \vnt )$.  Therefore, by using the method described in \cite{1}, one can determine
  the corresponding quadratic form $Q$ on the torsion group
\be
Q(\gamma = n h ) = n^2 r/p \; .
\label{7.1}
\ee
Consequently the path-integral partition function (\ref{5.55}) is given by
\be
Z_k (L_{p/r}) =  \sum_{n=0}^{p-1}\,
\exp \left ( {2\pi i k r \over  p } \, n^2  \right ) \; .
\label{7.2}
\ee
Let us now consider the link $L = C_1 \cup C_2 \subset L_{p/r}$ which, in a surgery presentation of $ L_{p/r}$, is shown in Figure~7.1(a).

  \vskip 0.8 truecm
\centerline {\includegraphics[width=3.5in]{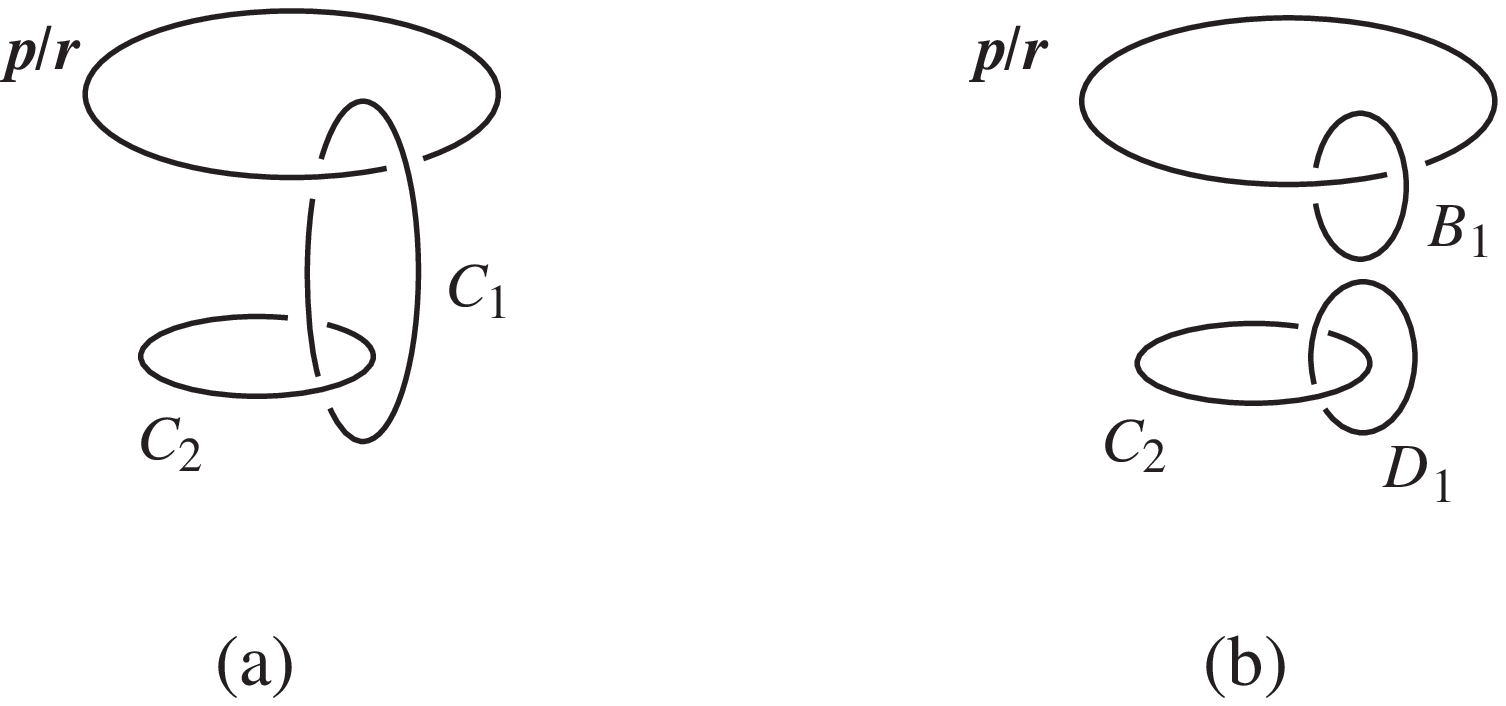}}
\vskip 0.3 truecm
\centerline {{Figure 7.1.} {Links in the lens space $L_{p/r}$.}}
\vskip 0.4 truecm

\noindent   Let $q_1$ and $q_2$ be the charges associated with $C_1$ and $C_2$ respectively, and suppose that the orientations and the framings of the link $L$ are specified by the following data
\be
\ell k (C_1 , C_2 ) \Big |_{S^3} = 1 \quad , \quad
\ell k (C_1 , C_{1 {\rm f}} ) \Big |_{S^3} = f_1 \quad , \quad
\ell k (C_2 , C_{2 {\rm f}} ) \Big |_{S^3} = f_2 \; ,
\label{7.3}
\ee
where the linking numbers refer to the sphere of the surgery presentation.   The component $C_1$ ---whose homology class is nontrivial---  can be understood as the band connected sum of the two knots $B_1 $ and $D_1$ shown in Figure~7.1(b); therefore
\be
 \langle\!  \langle  W_{C_2} W_{C_1} \rangle \! \rangle \Big |_{L_{p/r}} =    \langle\!  \langle W_{C_2} W_{D_1 } W_{B_1} \rangle \! \rangle \Big |_{L_{p/r}} \; .
 \label{7.4}
 \ee
 Note that the link $  C_2 \cup D_1$ belongs to the interior of a 3-ball in $L_{p/r}$ and then its contribution to the expectation value factorizes and coincides with the contribution in the sphere $S^3$,
\be
  \langle\!  \langle W_{C_2} W_{D_1 } W_{B_1} \rangle \! \rangle \Big |_{L_{p/r}} =     \langle\!  \langle W_{C_2} W_{D_1 }  \rangle \! \rangle \Big |_{S^3}  \, \langle\!  \langle  W_{B_1} \rangle \! \rangle \Big |_{L_{p/r}}   \; .
 \label{7.5}
 \ee
 The knots $B_1$ and $D_1$ have charge  $q_1$ and their orientation is determined by the orientation of $C_1$; let us now consider the framings. It is convenient to choose the framing of $B_1$ to be trivial with respect to the sphere $S^3$ of the surgery presentation, i.e. $\ell k (B_1 , B_{1 {\rm f}} )  |_{S^3}= 0 $.  In fact in this case one has $\ell k (D_1 , D_{1{\rm f}}) |_{S^3} = \ell k (C_1 , C_{1{\rm f}}) |_{S^3} = f_1$, and therefore  the link $ D_1 \cup C_2 $ would be ambient isotopic with  the original link $ C_1 \cup C_2 $   in $S^3$ if  the surgery instructions are neglected,  that is
 \be
 \langle\!  \langle W_{C_2} W_{D_1 }  \rangle \! \rangle \Big |_{S^3}   =   \langle W_{C_2} W_{D_1 }  \rangle \Big |_{S^3} =   \langle W_{C_2} W_{C_1 }  \rangle \Big |_{S^3}   \; .
 \label{7.6}
 \ee
  So one finds
 \bea
 \langle\!  \langle W_{C_1} W_{C_2} \rangle \! \rangle \Big |_{L_{p/r}} &=&   \langle W_{C_1} W_{C_2 }  \rangle \Big |_{S^3} \, \langle\!  \langle  W_{B_1} \rangle \! \rangle \Big |_{L_{p/r}} \nonumber \\
 &=& \exp \left [ - (2 \pi i / 4k) \left ( f_1 q_1^2 + f_2 q_2^2 + 2 q_1 q_2 \right ) \right ]  \, \langle\!  \langle  W_{B_1} \rangle \! \rangle \Big |_{L_{p/r}} \; .
 \label{7.7}
\eea
The knot $B_1$ has charge $q_1$ and has trivial framing with respect to the sphere $S^3$ of the surgery presentation. Since  $B_1$  is a representative of the generator $h \in T(L_{p/r})$, the knot $B_1^p$ ---which denotes the $p$-covering of the knot $B_1$, as in \S~5.4--- is homologically trivial and,   since it is ambient isotopic with $f^* (\mu)$,  its linking number in $L_{p/r}$  is given by
\be
\ell k (B_1^p , B^p_{1 {\rm f}}) \Big |_{L_{p/r}}  = - pr \; .
\label{7.8}
\ee
Therefore the perturbative component (\ref{5.48}) of $ \langle\!  \langle  W_{B_1} \rangle \! \rangle \big |_{L_{p/r}}$ is given by
\be
{\int  D  \omega \, e^{iS[  \omega ] } e^{2 \pi i q_1 \oint_{B_1}   \omega }\over \int D \omega \, e^{i S[\omega ]}}  = \exp \left [ - {2 \pi i \over 4 k } \,      q_1^2  \left ( { - p r \over  p^2}  \right ) \right ]    \; .
\label{7.9}
\ee
Let us now consider the nonperturbative component  (\ref{5.40}) of $ \langle\!  \langle  W_{B_1} \rangle \! \rangle \big |_{L_{p/r}}$. The canonical origin $A^0_\gamma$ of the $H^1_D(L_{p/r})$ fibre over $\gamma = n h \in T(L_{p/r})$ can be written as
\be
A^0_\gamma = n A^0 \; ,
\label{7.10}
\ee
where $A^0$ is the gauge orbit of a flat connection which corresponds to the generator of the torsion group. In agreement with equation (\ref{7.1}), the value of the CS action $S[A^0]$ reads
\be
e^{2 \pi i k \int A^0 * A^0 } = e^{2 \pi i k \, r / p }\; ,
\label{7.11}
\ee
and the holonomy along $B_1$ ---which can be evaluated by using the methods described in \cite{1,9}--- is given  by
\be
e^{i q_1 \oint_{B_1} A^0 } = e^{2 \pi i  q_1 r / p }\; .
\label{7.12}
\ee
Consequently, the nonperturbative component
of $ \langle\!  \langle  W_{B_1} \rangle \! \rangle \big |_{L_{p/r}}$ takes the form
\be
\sum_{n =0}^{p -1}  e^{2 \pi i k  n^2 \int A^0 *   A^0 }  e^{2 \pi i  n \oint_{B_1}  A^0 }  =
\sum_{n=0}^{p-1}  \exp \left [ {{2 \pi i  r  \over  p} ( k n^2 +  n q_1 ) } \right ]  \; .
\label{7.13}
\ee
By combining expressions (\ref{7.9}) with (\ref{7.13}) one finds
\be
\langle\!  \langle  W_{B_1} \rangle \! \rangle \Big |_{L_{p/r}} =  \sum_{n=0}^{p-1}\,
\exp \left [ {2\pi i k r \over  p } (n + q_1 / 2k)^2  \right ] \; .
\label{7.14}
\ee
Expression (\ref{7.14}) is defined for values of $q_1$ that belong to the residue classes of integers mod~$2k$, as it should be. A second check of equation (\ref{7.14})  can be   obtained by putting $q_1=p$. Indeed the knot $B_1$ with charge $q_1=p$ is equivalent to the knot $B_1^p$ (the $p$-covering of $B_1$) with unit charge. Since $B_1^p$  belongs to a 3-ball in $L_{p/r}$ and has self-linking number shown in equation (\ref{7.8}), expression (\ref{7.14})  should be equal to the expectation value in $S^3$ of the unknot (with charge $=1$) with self-linking number $-pr$ multiplied by the partition function $Z_k (L_{p/r})$. And in fact, when $q_1=p$, expression (\ref{7.14})  becomes
\be
 \langle\!  \langle  W_{B_1} \rangle \! \rangle \Big |_{L_{p/r} \; , \; q_1 = p } =  e^{- (2 \pi i / 4 k) (- pr) }  \sum_{n=0}^{p-1}\;
\exp \left [ {2\pi i k r \over  p } n^2  \right ] \; ,
\label{7.15}
\ee
which  is in agreement with equation (\ref{7.2}).

\section {Conclusions}

The successful achievement of the functional integration in the $U(1)$ Chern-Simons theory ---which is defined in a general 3-manifold $M$---  illuminates some open problems that one encounters in gauge quantum field theories when the space (or spacetime) manifold has nontrivial topology.

The group of local $U(1)$ gauge transformations is extended with respect to the  $S^3$ case and is described by the set of closed 1-forms with integral periods. The  path-integral is defined in the space of the gauge orbits of the connections which belong to the various  inequivalent $U(1)$ principal  bundles over $M$.  The integration in each sector of the configuration space takes the form of a standard functional integration over 1-forms  (modulo gauge transformations) in the presence of appropriate background connections, the sum over all the inequivalent principal bundles is given by a sum over the backgrounds. When the manifold $M$ has nontrivial topology the central issue is the choice of the normalization of the path-integral. With gauge group $U(1)$, the functional integration over the  gauge orbits of the connections which belong to  the trivial $U(1)$ principal bundle over $M$ represents the  canonical normalization, which generalises the ordinary path-integral normalization in the case of the sphere $S^3$ and permits to give a meaningful definition of the partition function in a general manifold $M$.

A few technical aspects of the actual nonperturbative computation of the $U(1)$ Chern-Simons path-integral are based on the particular form of the action and of the observables. In the computation of gauge-invariant observables, any gauge-fixing procedure can be avoided and in the presence of zero modes ---where standard perturbation theory cannot be used because the fields propagator does not exist--- one can still carry out the functional integration; indeed the amplitudes of the zero modes take values in a compact space (because local gauge transformations are described by closed 1-forms with integral periods) and the corresponding path-integral is finite. The topology of the manifold $M$ is revealed by the gauge orbits of flat connections, which dominate the functional integration in a real way (not only in the semiclassical approximation).

The  path-integral invariants are  related  with the Reshetikhin-Turaev surgery invariants by  a multiplicative factor that, according to the Deloup-Turaev reciprocity formula, only   depends on the torsion numbers and on the first Betti number of the manifold $M$.

 The $U(1)$ Chern-Simons gauge field theory and its description in terms of the Deligne-Beilinson formalism admit a  natural extension \cite{28}   to the case of closed $(4n+3)$-manifolds.
Also in these higher-dimensional models, the computation of the path-integral invariants ---like the partition function and the expectation value of the gauge-invariant holonomies---  can be achieved by using the methods illustrated in the previous sections. Similarly to the 3-dimensional formula (\ref{5.54}), the path-integral invariants depend on the higher-dimensional linking numbers and on a linking quadratic form  on the appropriate  torsion group. Note that one can always transform an expression of the type (\ref{5.54})  and rewrite it \cite{29}  as a suitable combination of invariants ---functions of linking numbers--- computed in the sphere $S^{(4n+3)}$.  Let us now consider the higher-dimensional surgery invariants. A generic $(4n+3)$-manifold with $n \geq 1$ is not necessarily cobordant with the sphere and then it may not admit a Dehn surgery presentation in $S^{(4n+3)}$; thus a general construction of surgery invariants, which are the analogue of the 3-dimensional Reshetikhin-Turaev invariants, appears  rather problematic.  Nevertheless, the possibility of finding an appropriate  combination of abelian invariants of the sphere  $S^{(4n+3)}$ which represents an invariant of a  $(4n+3)$-manifold ---even in the absence of a general Dehn surgery presentation of the manifold in the sphere $S^{(4n+3)}$--- has been recognised by  Deloup in Ref.\cite{30}. Thus, the path-integral invariants of the $U(1)$ Chern-Simons field theory defined in a $(4n+3)$-manifold give an explicit realisation of the Deloup  prediction.

We have shown that the $U(1)$ Chern-Simons path-integral invariants can be written as sums of exponentials of specific linking numbers. Now, appropriate combinations of linking pairings can also be used to define new topological  invariants; one (cubic) example of this type has been produced by Lescop \cite{31}.

In addition to the path-integral method of quantum field theory that has been discussed in the present article, one can consider different and mathematical approaches  to  the $U(1)$ Chern-Simons theory as presented,  for instance,  in the papers  \cite{32}--\cite{51}.

\begin{appendix}

\section {Fundamentals of abelian gauge symmetry}

This appendix contains some basic definitions concerning abelian gauge theories in a general  topologically nontrivial manifold, and includes  the used conventions of the Deligne-Beilinson formalism \cite{11,12,52,53,54,55}.

\subsection{Good cover and polyhedral decomposition}
  It is convenient to provide the closed oriented 3-manifold $M$  with a  good cover ${\cal U}$, which is given by  a collection of open sets ${\cal U}_a$ of $M$ such that $\bigcup_a {\cal U}_a = M$; moreover  each non-empty open set ${\cal U}_{a_1 a_2 \cdots a_m} := {\cal U}_{a_1} \cap {\cal U}_{a_2} \cap \cdots \cap {\cal U}_{a_m} $ is homeomorphic to $\hbox{\d R}^3$ and hence it is contractible. The index of ${\cal U}_{a_1 a_2 \cdots a_m}$ is refereed as the \v Cech index of this intersection and the integer $m$ as its  \v Cech degree. For later convenience, we only consider intersections ${\cal U}_{a_0 a_1 \cdots a_m}$ whose {\v Cech} indexes are ordered according to $a_0 < a_1 < \cdots < a_m$. We say that $\cal U$ is an  ordered  good cover. Furthermore, since $M$ is compact, we can assume that $\cal U$ is   finite.

Poincar\'e lemma applies in any intersection of the finite ordered good cover $\cal U$; this means  $d \omega = 0 \Leftrightarrow \omega = d \chi$ in any ${\cal U}_{a_0 a_1 \cdots a_m}$. Strictly speaking Poincar\'e lemma only holds for $p$-forms with $p > 0$. If $f$ is a $0$-form ({\it i.e.}, a  function) defined in ${\cal U}_{a_0 a_1 \cdots a_m}$ such that $df = 0$, then $f = \hbox{constant}$ in ${\cal U}_{a_0 a_1 \cdots a_m}$; one then extends the de Rham exterior derivative $d$ by the canonical injection of numbers into (constant) functions, denoted $d_{-1}$, so that Poincar\'e lemma also extends to functions. Obviously $d d_{-1} = 0$, and hence the fundamental property $d^2 = 0$ is still fulfilled.

The space of (singular oriented) $p$-cycles in $M$ is denoted by ${\cal Z}_p(M)$, $0 \leq p \leq 3$. The complete mathematical description of cycles in $M$ is not required so one can see $p$-cycles in $M$ as closed $p$-dimensional submanifolds of $M$, and $p$-chains as $p$-dimensional submanifolds whose boundaries are a closed $(p-1)$-dimensional submanifolds, the boundary operator bing denoted by $b$. For instance, a  knot is a smooth mapping $C: S^1 \rightarrow M$ such that $C(S^1)$ is homeomorphic to $S^1$. The space ${\cal Z}_p(M)$ is a ${\mathbb Z}$-module: any  integral combination of $p$-cycles is a $p$-cycle. For instance, $- C$ amounts to reversing the orientation of the knot $C$, whereas ---at the classical level--- $n C$ amounts to travel $n$ times along the knot $C$. The integer $n$ is also refereed as the charge of the colored knot $n C$.

In order to address integration in $M$, we shall use the concept of  polyhedral decomposition.    A polyhedral decomposition subordinate to a good cover $\cal U$ of a $p$-cycle $N$ of $M$ is defined as follows: first, decompose $N$ into $p$-dimensional components $N_{p}^{a_0}$ such that
\be
N = \sum_{a_0} N_{p}^{a_0} \quad , \quad  N_{p}^{a_0} \subset {\cal U}_{a_0} \, .
\label{A.1}
\ee
To prevent overcounting one has to select which $N_{p}^{a_0}$ is nonvanishing and really does contribute to the previous sum and which does not. In other words, one associates to each ${\cal U}_{a}$ a component $N_{p}^{a}$ of $N$ and some of these components may be zero. Note that the finiteness of $\cal U$ ensures that the sum is always finite.

The boundaries $b N_{p}^{a_0}$ form a collection of $(p-1)$-submanifolds each of which is decomposed on its turn into $(p-1)$-dimensional pieces $N_{p-1}^{a_0 a_1}$ according to
\be
b N_{p}^{a_0} = \sum_{a_1}  (N_{p-1}^{a_1 a_0} - N_{p-1}^{a_0 a_1}) \quad , \quad N_{p-1}^{a_0 a_1} \subset {\cal U}_{a_0 a_1} \, .
\label{A.2}
\ee
As in the previous step, one has to select which $N_{p-1}^{a_0 a_1}$ really contributes to the decomposition, putting all the others equal to zero. Furthermore, the ordering of $\cal U$ induces an ordering in the indices of the components $N_{p-1}^{a_0 a_1}$. For instance, suppose that in the decomposition of $b N_{p}^{a_0}$ the component  $N_{p-1}^{a_0 a_1}$ is nonvanishing and  $a_0 < a_1$,  in this case,   it is the term $- N_{p-1}^{a_0 a_1}$ (and not $+ N_{p-1}^{a_1 a_0}$) which really contributes to the sum (\ref{A.2}). Whereas in the decomposition of $b N_{p}^{a_1}$ it is the term $+ N_{p-1}^{a_0 a_1}$ which contributes to the sum.

The ordered components $N_{p-1}^{a_0 a_1}$ also have boundaries, and hence the decomposition is continued according to:
\be
b N_{p-1}^{a_0 a_1} = \sum_{a_2} (N_{p-2}^{a_2 a_0 a_1} - N_{p-2}^{a_0 a_2 a_1} + N_{p-2}^{a_0 a_1 a_2 }) \quad  , \quad N_{p-2}^{a_0 a_1 a_2 } \subset {\cal U}_{a_0 a_1 a_2} \, .
\label{A.3}
\ee
If $N_{p-2}^{a_0 a_1 a_2}$ is nonvanishing in the decomposition and if $a_0 < a_1 < a_2$,  then it is the term $+ N_{p-2}^{a_0 a_1 a_2}$ which contributes to the sum (\ref{A.3}), whereas the term  $- N_{p-2}^{a_0 a_1 a_2}$  really contributes to the sum for the decomposition of $b N_{p}^{a_0 a_2}$  and the term $+ N_{p-1}^{a_0 a_1 a_2}$ contributes to the sum for  $b N_{p}^{a_1 a_2}$.

The construction illustrated above is iterated, thus generating a collection of $(p-3)$-dimensional submanifolds $N_{p-2}^{a_0 a_1 a_2 a_3} \subset {\cal U}_{a_0 a_1 a_2 a_3}$ such that
\be
b N_{p-2}^{a_0 a_1 a_2} = \sum_{a_3} (N_{p-3}^{a_3 a_0 a_1 a_2} - N_{p-3}^{a_0 a_3 a_1 a_2} + N_{p-3}^{a_0 a_1 a_3 a_2} - N_{p-3}^{a_0 a_1 a_2 a_3}) \, ,
\label{A.4}
\ee
with the same ordering convention as before and the same selection principle on the contributing components. This is the last step of the decomposition process since the submanifolds of a 3-manifold  are at most of dimension $3$ so that $N_{p-3}^{a_0 a_1 a_3 a_2}$ are points when $p = 3$. The procedure stops at the first stage when $p = 0$, at the second stage when $p = 1$ and at the third stage when $p = 2$. In other words, a polyhedral decomposition subordinate to $\cal U$ gives the following possible sequences:
\be
\begin{array}{l@{\; : \;} l}
M & \{ M_{3}^{a_0} , S_{2}^{a_0 a_1} , l_{1}^{a_0 a_1 a_2} , x_{0}^{a_0 a_1 a_2 a_3} \}  \\
\Sigma & \{ \Sigma_{2}^{a_0} , l_{1}^{a_0 a_1} , x_{0}^{a_0 a_1 a_2} \}   \\
C & \{ C_{1}^{a_0} , x_{0}^{a_0 a_1} \} \\
X & \{ x_{0}^{a_0} \}   \, ,
\end{array}
\label{A.5}
\ee
where $\Sigma$ is a closed surface in $M$, $C$ a knot in $M$ and $X$ a collection of points in $M$. One can check that for a fixed $\cal U$ not all $p$-cycles of $M$ admits a polyhedral decomposition subordinate to $\cal U$. However, it is always possible to find an ordered good cover of $M$ with respect to which a given $p$-cycle admits a polyhedral decomposition. So we will always assume that a well-adapted finite ordered cover has been chosen when dealing with a polyhedral decomposition.

\vskip 0.5 truecm
\centerline {\includegraphics[width=2.30 in]{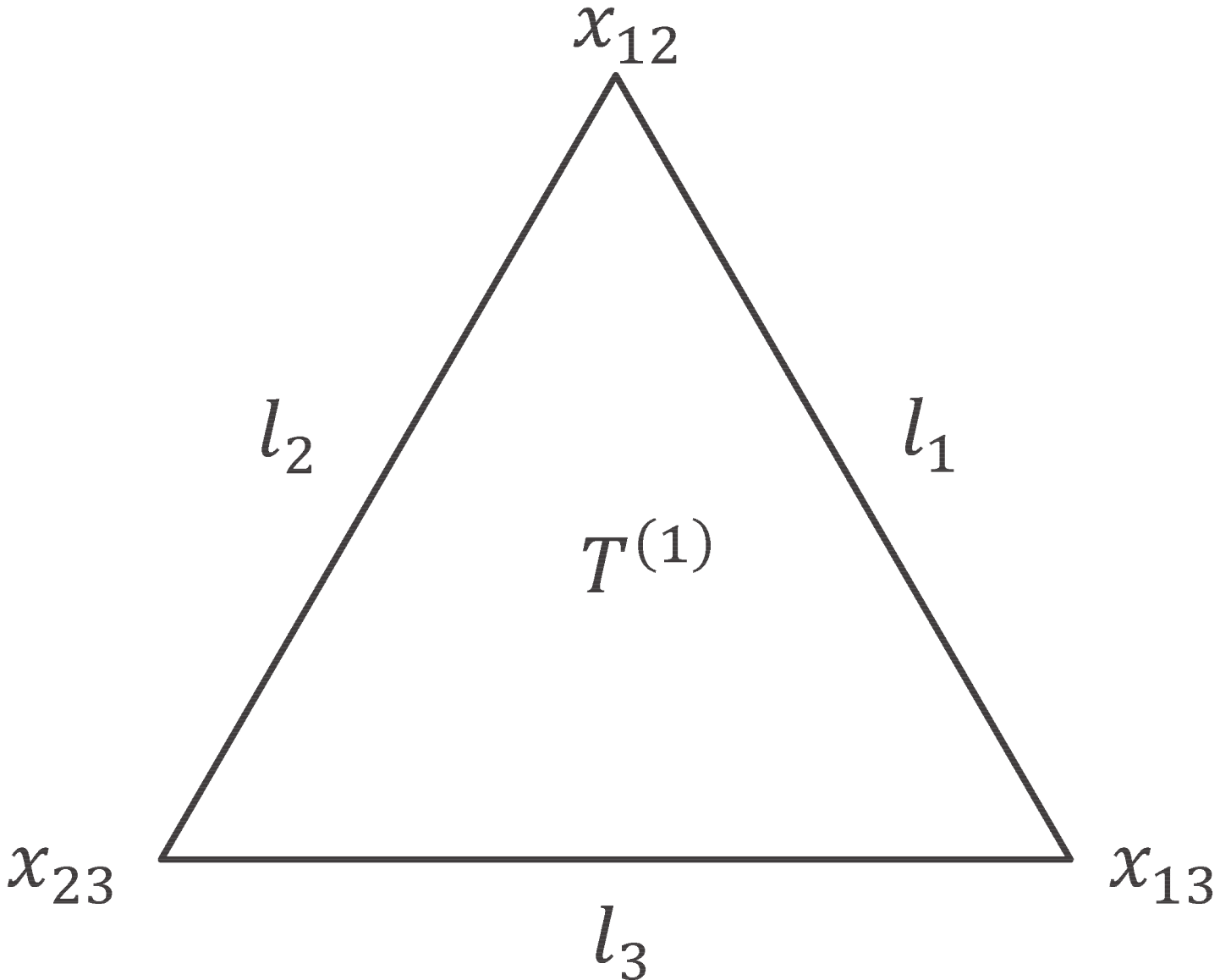}}
\vskip 0.1 truecm
\centerline {{Figure A.1.} {A non-oriented polyhedral decomposition of a triangular knot.}}

\vskip 0.3 truecm

As a first example, let us consider an oriented triangle $T^{(1)}$ as a $1$-cycle in $M$. This triangle is the sum of its three oriented edges, $l_1$, $l_2$ and $l_3$, as depicted in Figure~A.1.  Hence: $T^{(1)} = l_1 + l_2 + l_3$. The ends of these edges are made of three points such that $b l_1 = (x_{21} - x_{12}) + (x_{31} - x_{13}) = - x_{12} - x_{13}$, $b l_2 = (x_{12} - x_{21}) + (x_{32} - x_{23}) = x_{12} - x_{23}$ and $b l_3 = (x_{13} - x_{31}) + (x_{23} - x_{32}) = x_{13} + x_{12}$. Then:  $b T^{(1)}  = b l_1 + b l_2 + b l_3 = - x_{12} - x_{13} + x_{12} - x_{23} + x_{13} + x_{12} = 0$, as it has to be.

As a second example, let us consider an oriented tetrahedron $T^{(2)}$ as a $2$-cycle in $M$. It is made of 4 oriented triangular faces, $\Delta_a$ ($a = 1 , \cdots , 4$), bond to each other on their edges, as depicted in Figure A.2.

\vskip 0.5 truecm
\centerline {\includegraphics[width=2.30 in]{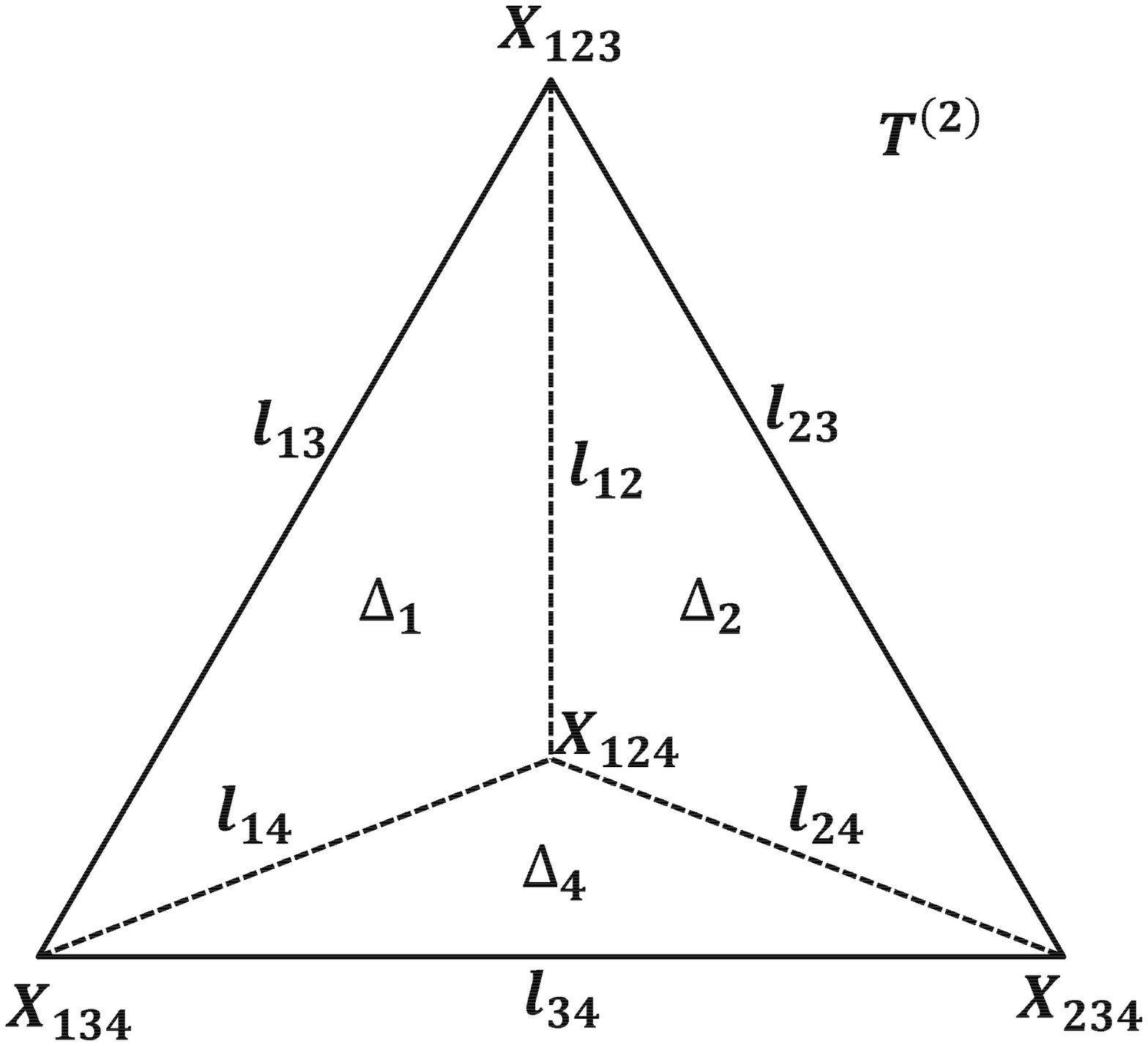}}
\vskip 0.1 truecm
\centerline {{Figure A.2.} {A non oriented polyhedral decomposition of a tetrahedral surface.}}
\centerline {{The face $\Delta_3$ is the front face and has been omitted.}}

\vskip 0.3 truecm

\noindent Note that $T^{(1)}$ is a topological representative of a $2$-sphere. Then one has:
$$
\begin{array}{l@{\; =\; } l}
b \Delta_1 & (l_{21} - l_{12}) + (l_{31} - l_{13}) + (l_{41} - l_{14}) = - l_{12} - l_{13} - l_{14} \, , \\
b \Delta_2 & (l_{12} - l_{21}) + (l_{32} - l_{23}) + (l_{42} - l_{24}) = + l_{12} - l_{23} - l_{24} \, , \\
b \Delta_3 & (l_{13} - l_{31}) + (l_{23} - l_{32}) + (l_{43} - l_{34}) = + l_{13} + l_{23} - l_{34} \, , \\
b \Delta_4 & (l_{14} - l_{41}) + (l_{24} - l_{42}) + (l_{34} - l_{43}) = + l_{14} + l_{24} + l_{34} \, ,
\end{array}
$$
with $b P = \sum_a b \Delta_a = 0$ as expected. And finally:
$$
\begin{array}{l@{\; =\; } l}
b l_{12} & (x_{312} - x_{132} + x_{123}) + (x_{412} - x_{142} + x_{124}) = + x_{123} + x_{124} \, , \\
b l_{13} & (x_{213} - x_{123} + x_{132}) + (x_{413} - x_{143} + x_{134}) = - x_{123} + x_{134} \, , \\
b l_{14} & (x_{214} - x_{124} + x_{142}) + (x_{314} - x_{134} + x_{143}) = - x_{124} - x_{134} \, , \\
b l_{23} & (x_{123} - x_{213} + x_{231}) + (x_{423} - x_{243} + x_{234}) = + x_{123} + x_{234} \, , \\
b l_{24} & (x_{124} - x_{214} + x_{241}) + (x_{324} - x_{234} + x_{243}) = + x_{124} - x_{234} \, , \\
b l_{34} & (x_{134} - x_{314} + x_{341}) + (x_{234} - x_{324} + x_{342}) = + x_{134} + x_{234} \, .
\end{array}
$$
One can check that: $b l_{12} + b l_{13} + b l_{14} = 0$, $b l_{12} - b l_{23} - b l_{24} = 0$, $b l_{13} + b l_{23} - b l_{34} = 0$ and $b l_{12} + b l_{24} + b l_{34} = 0$, which is consistent with  $b^2 = 0$. The 4 points defining the vertices of $T^{(2)}$ inherit an orientation from the previous equations.

If in these examples the indices of the various faces, edges and vertices are referring to the intersections of a good cover $\cal U$ of $M$, then the two decomposition are subordinate to $\cal U$.

\subsection {Gauge orbits}    A $U(1)$ gauge field $\cal A$ on $M$ is defined by a triplet of local variables
\be
{\cal A} = \{ v_a, \lambda_{a b} , n_{a b c} \} \; ,
\label{A.6}
\ee
where the ``vector" fields $v_a$'s are 1-forms in the open sets ${\cal U}_a$, whereas the scalar fields $\lambda_{a b}$'s are 0-forms (functions) in the intersections ${\cal U}_{a b}$, and the $n_{abc}$'s are integers defined in the intersections ${\cal U}_{a b c}$ such that the $d_{-1} n_{abc}$'s are constant scalar fields.
The various variables appearing in expression (\ref{A.6}) are ordered with respect to the values of their degrees; more precisely, when the de Rham degree ({\it i.e.}, the form degree) lowers, then the \v Cech degree increases.
The functions $\lambda_{ab}$ specify how the 1-forms $v_a $ and $v_b $ are related in the intersection ${\cal U}_{a b}$:
\be
(\delta v)_{ab} := v_b - v_a = d \lambda_{a b} \; .
\label{A.7}
\ee
These relations encode the gauge ambiguity of the local vector fields $v_a$. Similarly, the integers $n_{abc}$ describe the behavior of the 0-forms $\lambda_{ab}$, $\lambda_{ac}$ and $\lambda_{bc}$ in the intersections ${\cal U}_{a b c}$,
\be
(\delta \lambda )_{abc} := \lambda_{b c} - \lambda_{a c} + \lambda_{a b} = d_{-1} n_{a b c} = n_{a b c} \; ,
\label{A.8}
\ee
so that:
\be
(\delta n)_{abcd} := n_{b c d} - n_{a c d} + n_{a b d} - n_{a b c} = 0 \; ,
\label{A.9}
\ee
is tautologically fulfilled in the intersections ${\cal U}_{a b c d}$. This last equation means that the collection $\{ n_{a b c} \}$ is an  integral \v Cech cocycle  of $\cal U$. On the other hand, equation (\ref{A.7}) implies that, in each intersection  ${\cal U}_{a b}$, the local $2$-form $ d v_a$ and $ d v_b$  satisfy $dv_b - dv_a = d (v_b - v_a) = d (d \lambda_{ab}) = 0$. Hence, the collection $\{ dv_a \}$ can be identified with the set of local representatives of a closed $2$-form $F_{\cal A}$. This form is precisely the curvature of $\cal A$. Finally equations (\ref{A.7})--(\ref{A.9}) imply that $F_{\cal A}$ has integral periods; that is to say,  for any closed surface $\Sigma$ in $M$ one has
\be
\int _\Sigma F_{\cal A} = n  \in {\mathbb Z} \; ,
\label{A.10}
\ee
which is equivalent to
$$
\exp \left ( 2\pi i \int _\Sigma F_{\cal A} \right ) =1 \; .
$$
Indeed, if $\{ \Sigma_a, l_{ab},x_{abc} \}$ is a polyhedral decomposition of $\Sigma$ then
$$
\int _\Sigma F_{\cal A} = \sum_a \int _{\Sigma_a} d v_a = \sum_a \int _{b \Sigma_a} v_a = \sum_{a,b} \int _{l_{ba} - l_{ab}} v_a = \sum_{a,b}  \int _{l_{ab}} (v_b - v_a) \, .
$$
Equation (\ref{A.7}) then gives

\bea
\int _\Sigma F_{\cal A} &=& \sum_{a,b}  \int _{l_{ab}} d \lambda_{ab} = \sum_{a,b}  \int _{b l_{ab}} \lambda_{ab} = \sum_{a,b,c}  \int _{x_{cab} - x_{acb} + x_{abc}} \mkern-90mu \lambda_{ab} \nonumber \\
&=& \sum_{a,b,c}  \int _{x_{abc}}  \mkern-10mu (\lambda_{bc} - \lambda_{ac} + \lambda_{ab}) \, , \nonumber
\eea
and from equation (\ref{A.8}) one gets
$$
\int _\Sigma F_{\cal A} = \sum_{a,b,c}  \int _{x_{abc}} \mkern-10mu d_{-1} n_{a b c} := \sum_{a,b,c} (d_{-1} n_{a b c})(x_{abc}) \in {\mathbb Z} \, ,
$$
because each $(d_{-1} n_{a b c})$ is by construction a ${\mathbb Z}$-valued function in ${\cal U}_{ab}$.

A $U(1)$ connection on $M$ can also be interpreted as the image on the manifold   of a connection on a $U(1)$ principal bundle over $M$. The bundle transition functions $g_{a b } : {\cal U}_a \cap {\cal U}_b \rightarrow U(1)$ are given by
\be
g_{a b } =  e^{2 \pi i \lambda_{a b }} \; .
\label{A.11}
\ee
Equation (\ref{A.8}) ensures that, in the intersections ${\cal U}_a \cap \, {\cal U}_b \cap \, {\cal U}_c $, the consistency condition
$$
g_{a b } \, g_{ b c } \, g_{ c a } = 1
$$
is satisfied. Thus the \v Cech-de Rham presentation (\ref{A.6}) of the connection $\cal A$ actually specifies a $U(1)$ principal bundle  with connection. In our notations the so-called connection 1-form is locally represented by $ 2 \pi v_a $ and a   gauge transformation ---associated with the  group element $ g_a = \exp (2 \pi i \xi_a  ) $ in the open set ${\cal U}_a$--- takes the form
\be
\left \{ \begin{array} {l@{\; \rightarrow \;} l}
 2 \pi v_a &  e^{-2\pi i \xi_a  } 2 \pi v_a e^{2\pi i \xi_a } - ie^{-2\pi i \xi_a } d \, e^{2\pi i \xi_a } = 2 \pi ( v_a + d \xi_a )\; , \\
 g_{a b } & e^{-2\pi i \xi_a  } g_{a b} e^{2\pi i \xi_b }  \quad , \quad n_{a b c} {\rightarrow }n_{a b c } \; ,
 \end{array} \right.
\label{A.12}
\ee
where each function $ \xi_a $ is defined in $ {\cal U}_a $. Note that, on the components of $\cal A$, a general gauge transformation reads
\be
\left \{ \begin{array}  {l@{\; \rightarrow \;} l}
 v_a & v_a + d\xi_a \; , \\
  \lambda_{ab} & \lambda_{ab} + \xi_b - \xi_a - m_{ab} = \lambda_{a b} + (\delta \xi )_{ab} - m_{ab} \; , \\
  n_{abc} & n_{abc} - m_{bc} + m_{ac} - m_{ab} = n_{abc} - (\delta m)_{abc} \; ,
  \end{array}  \right.
\label{A.13}
\ee
where the free parameters $m_{ab}$'s (with $m_{ba}= - m_{ab}$) are integers which are defined in the intersections ${\cal U}_a \cap  {\cal U}_b $. These integers do not appear in equation (\ref{A.12}) because the restricted transformation
\be
\left \{
\begin{array} {l@{\; \rightarrow \;} l}
v_a & v_a \; , \\
 \lambda_{ab} & \lambda_{ab} - m_{ab} \; , \\
 n_{abc} & n_{abc} - (\delta m)_{abc}  \; , \end{array}  \right.
\label{A.14}
\ee
 preserves conditions (\ref{A.7}) and (\ref{A.8}) and does not modify the bundle transition functions (\ref{A.11}).

\subsection {Gauge holonomies}
Integrals of a $U(1)$ gauge field over 1-cycles  of $M$ (along oriented knots in $M$) are ${\mathbb R} / {\mathbb Z}$-valued and define the $U(1)$ holonomies of the gauge fields. More precisely, the holonomy of a $U(1)$ gauge field $\cal A $ on $M$ is a morphism $W : {\cal Z}_1(M) \rightarrow U(1)$, where ${\cal Z}_1(M)$ is the abelian group of 1-cycles of $M$. In the quantum field theory context, one really has to consider oriented and framed knots in $M$ because products of fields at the same point give rise to ambiguities in the mean values. In facts, this is precisely the reason why  the quantum expectation values of the knot holonomies  need to be defined for framed knots.  If the knot $C \subset M$ belongs to a single chart ${\cal U}_a$, the holonomy of the gauge field $\cal A$ along $C$ is defined by
$$
 W_C ({\cal A})  = e^{2 \pi i \oint_C {\cal A}} = e^{2 \pi i \oint_C v_a} \; .
$$
For a generic knot $C \subset M$, one first introduces a polyhedral decomposition $\{ C_a , x_{ab} \}$ of $C$ (subordinate to $\cal U$) and then consider the sum of integrals
\be
H_1 = \sum_{a} \int_{C_a} v_a \; .
\label{A.15}
\ee
If the collection of local $1$-forms $v_a$ defines a global $1$-form on $C$ then this sum reduces to the standard definition of the integral over $C$. Unfortunately, under a gauge transformation $v_a \rightarrow v_a + d \xi_a$, expression (\ref{A.15}) transforms as
\bea
H_1 \rightarrow H_1 + \sum_{a} \int_{ C_a} \mkern-10mu d \xi_a &=&  H_1 + \sum_{a} \int_{b C_a} \mkern-10mu \xi_a = H_1 + \sum_{a,b} \int_{x_{ba} - x_{ab}} \mkern-40mu \xi_a  \nonumber \\
&=& H_1 + \sum_{a,b} \int_{x_{ab}} \mkern-10mu ( \xi_b - \xi_a ) \, , \nonumber
\eea
where integration over points means evaluation. In order to eliminate the last term in this equation, one  can simply add to $H_1$ the term
\be
H_2 = - \sum_{a,b} \int_{X_{ab}} \mkern-15mu \lambda_{ab} \; ,
\label{A.16}
\ee
because a gauge transformation $v_a \rightarrow v_a + d \xi_a$ is accompanied by a transformation $\lambda_{ab} \rightarrow \lambda_{ab} + \xi_b - \xi_a$. Finally under the integral residual transformation $\lambda_{ab} \rightarrow \lambda_{ab} + d_{-1} m_{ab}$ the sum $H_1 + H_2$ transforms as
$$
H_1 + H_2 \rightarrow H_1 + H_2 + \sum_{a,b} \int_{X_{ab}} \mkern-10mu d_{-1} m_{ab} \, ,
$$
which does not modify the exponential $e^{2 i \pi (H_1 + H_2)}$. Hence, the reduction of $H_1 + H_2$ to ${\mathbb R} / {\mathbb Z}$ is a good candidate for defining $\oint_C \cal A$. With definition of  the holonomy of a $U(1)$ gauge field, gauge equivalent fields have the same holonomy along $C$, and any other polyhedral decomposition of $C$ (subordinate to $\cal U$) changes $H_1 + H_2$ by integral contributions thus defining also the same holonomy.

Hence, for any polyhedral decomposition $\{ C_a , x_{ab} \}$ of an oriented knot $C\subset M$, with color specified by the integer charge $q$, the holonomy $W_C ({\cal A} \, ; q )$ of the gauge field ${\cal A} = \{ v_a, \lambda_{a b} , n_{a b c} \}$ along $C$ is defined by
\be
W_C ( {\cal A}\, , q) = \exp \left ( 2 \pi i q \oint_C {\cal A }\right ) \equiv \exp \left [ 2 \pi i q \left ( \sum_{a} \int_{C_a} \mkern-10mu v_a - \sum_{a,b} \int_{x_{ab}} \mkern-15mu \lambda_{ab} \right ) \right ] \; .
\label{A.17}
\ee

\noindent When the knot $C$ is homologically trivial, $C = b \Sigma $, Stokes theorem implies
$$
W_C ( {\cal A} \, , q ) = e^{2 \pi i q \int_\Sigma F_{\cal A}} \; .
$$
When  the charge is quantized,  the holonomy (\ref{A.17}) represents a gauge invariant function of the gauge connection; therefore  $W_C$ is really  well defined for the DB classes in $H_D^1(M)$. In other words the holonomy of a DB class ({\it i.e.}, of a gauge orbit) $A$ along a knot $C$ is defined as the holonomy of any of its representative along $C$.

\subsection {The abelian Chern-Simons action}   Equations (\ref{A.7})--(\ref{A.9}) imply that, in general, $v_a$ is not the restriction in the open set ${\cal U}_a$ of a 1-form belonging to $\Omega^1(M)$. Similarly, the field combination $v_a \wedge d v_a$ is not necessarily the restriction  of a 3-form which is globally defined in $M$. Thus the  lagrangian of the CS gauge field theory with gauge group $U(1)$ ---which is is defined in a generic 3-manifold $M$--- cannot be written as $A\wedge dA$ where $A\in \Omega^1(M)$. In ${\mathbb R}^3$ the CS lagrangian takes the form  $A\wedge dA$ with $A\in \Omega^1({\mathbb R}^3)$ because any $U(1)$ principal bundle over $\hbox{\d R}^3$ is trivial (there is no nontrivial gluing procedure to implement).

The action of the $U(1)$ Chern-Simons gauge field theory in the 3-manifold $M$ is given by
\be
S[A] = 2 \pi k \int_M A * A \; ,
\label{A.18}
\ee
where $A$ represents the gauge orbit (or DB class) of a $U(1)$ gauge field on the manifold $M$, and  $A*A$ denotes the canonical DB product of $A \in H_D^1(M)$ with itself.  The $*$-product represents a pairing $H^1_D(M) \times H^1_D(M) \rightarrow H^3_D(M) \simeq {\mathbb R} / {\mathbb Z}$ that provides an appropriate generalization  of the $A\wedge d A$  expression ---which is well defined for 1-forms--- to the case of gauge orbits of $U(1)$ conenctions.

In order to produce the explicit expression of  $\int_M A * A $, let us  consider the gauge field (\ref{A.6}) and the collection of local 3-forms $v_a \wedge d v_a$ which are defined in the open sets ${\cal U}_a$. As in the case of the holonomy, it is convenient to use a polyhedral decomposition $\{ M^{a} , S^{ab} , l^{abc} , x^{abcd} \}$ of $M$ in order to try to define the desired integral. One first integrates the 3-forms $v_a \wedge d v_a$ on the 3-polyhedrons $M_a$ and sum over all the polyhedra
\be
I_1 = \sum_{a}  \int_{M_a}   v_a \wedge d v_a \; .
\label{A.19}
\ee
If the local fields $v_a$ actually define a $1$-form $v$ on $M$ then $I_1$ reduces to the well-defined standard form $\int_M v \wedge dv$. Under a gauge transformation $v_a \rightarrow v_a + d \xi_a$ one has $I_1 \rightarrow I_1 + \Delta I_1$ with
\bea
\Delta I_1 &=&  \sum_{a}  \int_{M_a}   d \xi_a \wedge d v_a =  \sum_{a}  \int_{M_a}   d (\xi_a \, d v_a) \nonumber \\  &=&  \sum_{a}  \int_{b M_a}   \xi^a \, d v_a = \sum_{a,b}  \int_{S_{ba} - S_{ab}}  \xi_a \, d v_a  \; . \nonumber
\eea
Since $dv_a$ is the restriction in ${\cal U}_a$ of a 2-form which is globally defined on $M$ (the curvature of $A$), one has
$$
\Delta I_1  =   \sum_{b,a}  \int_{S_{ab}}  (\xi_b - \xi_a)  d v_b  \; .
$$
Hence $\Delta I_1$ can take any value, thus preventing $e^{2\pi i I_1 }$ from being gauge invariant. In order to cancel $\Delta I_1$, one can introduce a new term $I_2$
\be
I_2 =  - \sum_{a,b}  \int_{S_{ab}}   \lambda_{ab}  d v_b \; .
\label{A.20}
\ee
Under the transformation $v_a \rightarrow v_a + d\xi_a$ and $\lambda_{ab} \rightarrow \lambda_{ab} + \xi_b - \xi_a$, one finds that $I_2 \rightarrow I_2 + \Delta I_2 $ with
$$
\Delta I_2 =  - \sum_{a,b}  \int_{S_{ab}}   (\xi_b - \xi_a)   d v_b \; .
$$
Therefore  the sum $I_1 + I_2$ is invariant under the  transformation $v_a \rightarrow v_a + d \xi_a $ and  $\lambda_{ab} \rightarrow  \lambda_{ab} + \xi_b - \xi_a$. Unfortunately $I_1 + I_2$ is not invariant under the integral residual transformations $\lambda_{ab} \rightarrow \lambda_{ab} + d_{-1} m_{ab}$, indeed under these transformations  one finds
$I_1 + I_2 \rightarrow I_1 + I_2 +  \widetilde  \Delta I_2$ with
\bea
 \widetilde \Delta I_2 &=&  \sum_{a,b} \int_{S_{ab}} m_{ab}  d v_b =  \sum_{a,b} \int_{S_{ab}} d (m_{ab}  v_b) \nonumber \\ &=&  -  \sum_{a,b}  \int_{b S_{ab}} m_{ab}  v_b
 =  \sum_{a,b,c} \int_{l_{cab} - l_{acb} + l_{abc}} \mkern-70mu m_{ab} v_b \nonumber \\
&=& \sum_{a,b,c} \int_{l_{abc}} \mkern-10mu ( m_{bc} v_c -  m_{ac} v_c + m_{ab} v_b) \nonumber \\
&=& \sum_{a,b,c} \int_{l_{abc}} \mkern-10mu (\delta m)_{abc} v_c + \sum_{a,b,c} \int_{l_{abc}} \mkern-10mu m_{ab} (\delta v)_{bc} \; . \nonumber
\eea
Hence the combination $I_1 + I_2$ is not gauge invariant. In order to cancel the first term in the right-hand side of $\widetilde \Delta I_2$, one can introduce the additional term $I_3$ given by
\be
I_3 =  \sum_{a,b,c} n_{abc}  \int_{l_{abc}}  v_c
 \; .
\label{A.21}
\ee
Under the transformation $\lambda_{ab} \rightarrow \lambda_{ab}  - d_{-1} m_{ab}$ and $n_{abc} \rightarrow n_{abc}  - (\delta m)_{abc}$ one has $I_3 \rightarrow I_3 + \widetilde  \Delta I_3$ with $\widetilde  \Delta I_3$ exactly compensating the first term of $\widetilde \Delta I_2$. The second term in $\widetilde \Delta I_2$ fulfills:
\bea
\sum_{a,b,c} \int_{l_{abc}} \mkern-10mu m_{ab} (\delta v)_{bc} &=& \sum_{a,b,c} \int_{l_{abc}} \mkern-10mu d ( m_{ab} \lambda_{bc}) = \sum_{a,b,c,d} \int_{x_{dabc} - x_{adbc} + x_{abdc} -x_{abcd}} \mkern-150mu  m_{ab} \lambda_{bc} \nonumber \\
&=& - \sum_{a,b,c,d} \int_{x_{abcd}} \mkern-10mu  (\delta m)_{abc} \lambda_{cd} -  \sum_{a,b,c,d} \int_{x_{abcd}} \mkern-10mu   m_{ab} (\delta \lambda)_{bcd} \nonumber \\
&=& - \sum_{a,b,c,d} \int_{x_{abcd}} \mkern-10mu  (\delta m)_{abc} \lambda_{cd} -  \sum_{a,b,c,d} \int_{x_{abcd}} \mkern-10mu   m_{ab} n_{bcd} \, . \nonumber
\eea
The last term of this expression is an integer so it does not break the gauge invariance of $e^{2 i \pi (I_1 + I_2 + I_3)}$ whereas the first term does. In order to compensate this contribution one can introduce a last contribution $I_4$
\be
I_4 = - \sum_{a,b,c,d}  \int_{x_{abcd}} n_{abc} \lambda_{cd} \, .
\label{A.22}
\ee
One can now verify that $e^{2 i \pi (I_1 + I_2 + I_3 + I_4)}$ is gauge invariant. Similarly, one can check that $e^{2 i \pi (I_1 + I_2 + I_3 + I_4)}$ is invariant under a change of polyhedral decomposition of $M$. Note that gauge invariance ensures that the result depends on the gauge orbit $A$ and not on its representatives.

Hence, for any polyhedral decomposition $\{ M^{a} , S^{ab} , l^{abc} , x^{abcd} \}$ of $M$ and any representative $\{ v_a, \lambda_{a b} , n_{a b c} \} $ of the class $A \in H_D^1(M)$, the quantum $U(1)$ Chern-Simons action, as a ${\mathbb R}/{\mathbb Z}$-valued quantity, is given by
\bea
S[A] \! \! &=& \! \! 2 \pi k \int_M A * A  =  2 \pi k \Biggl \{  \sum_{a}  \int_{M_a}  v_a \wedge d v_a -     \sum_{a,b}  \int_{S_{ab}} \lambda_{ab}  d v_b  \nonumber \\
&& {\hskip 2.3 truecm} -  \sum_{a,b,c}  n_{abc}  \int_{l_{abc}} v^c -  \sum_{a,b,c,d}  n^{abc} \int_{x_{abcd}} \lambda^{cd} \Biggr \} \, .
\label{A.23}
\eea

\vskip 0.4 truecm

\noindent Under a gauge transformation the action (\ref{A.23}) transforms as $S \rightarrow S + \Delta S$, where
\be
{\Delta S \over 2 \pi k }=  \sum_{x_{dcba}}  ( m^{bc} - m^{ac} + m^{ab} ) \, m^{cd}
\label{A.24}
\ee
takes integer values. Therefore the amplitude  $e^{i S[A]} $ has no ambiguities when the coupling constant $k$ is an integer. In facts, consistency of the formalism requires that, in addition to the values of the knot charges, the coupling constant $k$ also must have integer values. For a generic closed 3-manifold $M$, $S[A]$ is well defined (mod $\mathbb Z$) and reduces to the integral of  $A \wedge dA$ when $M$ is a homology sphere.

Expression (\ref{A.23}) of $\int A * A$ can also be used to define  the integral of $A * B$ on $M$ in terms of the representative field components of $A$ and $B$. In facts one can use the relation $2 \int A * B  = \int (A + B) * (A + B) - \int A * A - \int B * B$. It can be shown that any oriented knot $C$ in $M$ can be represented by a Deligne-Beilinson distributional class $\eta_C$ so that the integral of $A$ along $C$ coincides (modulo integers) with the integral of $A * \eta_C$ over $M$; this implies that the value of the exponential term $\exp \left ( 2 \pi i \int A * \eta_C \right ) $ is uniquely defined and verifies $\exp \left ( 2 \pi i \int A * \eta_C \right ) = \exp \left ( 2 \pi i \oint_C  A  \right )$, where $A * \eta_C$ is defined according to the previous construction of $A * A$.
\end{appendix}

\vskip 0.9 truecm

 \noindent {\bf Acknowledgments.}  We wish to thank R.~Benedetti and F.~Deloup for useful discussions and correspondence.

\vskip 0.7 truecm

\noindent{\bf References}

\medskip

\bibliographystyle{amsalpha}

\end{document}